\documentclass[10pt,journal,compsoc]{IEEEtran}
\usepackage{amsmath,amssymb,amsfonts,amsthm}
\usepackage{mathrsfs}
\usepackage{array}
\usepackage{booktabs}
\usepackage[caption=false,font=normalsize,labelfont=sf,textfont=sf]{subfig}
\usepackage{textcomp}
\usepackage{stfloats}
\usepackage{url}
\usepackage{verbatim}
\usepackage{multirow}
\usepackage{graphicx}
\usepackage{rotate}
\usepackage{amssymb}
\usepackage{cite}
\usepackage{xcolor}
\definecolor{color}{rgb}{0, 0, 1}
\definecolor{color1}{rgb}{1, 0, 0}
\usepackage{bm}
\usepackage[ruled,linesnumbered]{algorithm2e}

\usepackage{algpseudocode}
\makeatletter
\newcommand{\removelatexerror}{\let\@latex@error\@gobble}
\makeatother
\usepackage{mathtools}
\usepackage{epstopdf}
\usepackage{stfloats}

\begin{document}
	
	\title{Multi-objective Aerial Collaborative Secure Communication Optimization via Generative Diffusion Model-enabled Deep Reinforcement Learning}
	
	\author{
		Chuang Zhang,
		Geng~Sun\IEEEauthorrefmark{1},~\IEEEmembership{Senior Member,~IEEE,}
        Jiahui~Li,
	Qingqing~Wu,~\IEEEmembership{Senior Member,~IEEE,}
        Jiacheng~Wang,
        Dusit~Niyato,~\IEEEmembership{Fellow,~IEEE,}
		and Yuanwei~Liu,~\IEEEmembership{Fellow,~IEEE}
		\thanks{This study is supported in part by the National Natural Science Foundation of China (62172186, 62272194), and in part by the Science and Technology Development Plan Project of Jilin Province (20230201087GX). (Corresponding author: Geng Sun.)}
		\IEEEcompsocitemizethanks{
			\IEEEcompsocthanksitem Chuang Zhang and Jiahui Li are with the College of Computer Science and Technology, Jilin University, Changchun 130012, China, and also with the Key Laboratory of Symbolic Computation and Knowledge Engineering of Ministry of Education, Jilin University, Changchun 130012, China. E-mail: chuangzhang1999@gmail.com, lijiahui0803@foxmail.com.
                \IEEEcompsocthanksitem  Geng Sun is with the College of Computer Science and Technology, Jilin University, Changchun 130012, China, and also with the College of Computing and Data Science, Nanyang Technological University, Singapore 639798. E-mail: sungeng@jlu.edu.cn).
			\IEEEcompsocthanksitem Qingqing Wu is with the Department of Electronic Engineering, Shanghai Jiao Tong University, Shanghai, China. E-mail: qingqingwu@sjtu.edu.cn.
			\IEEEcompsocthanksitem Jiacheng Wang and Dusit Niyato are with the College of Computing and Data Science, Nanyang Technological University, Singapore 639798. E-mail: jiacheng.wang@ntu.edu.sg, dniyato@ntu.edu.sg. 
			\IEEEcompsocthanksitem  Yuanwei Liu is with the School of Electronic Engineering and Computer Science, Queen Mary University of London, London E1 4NS, U.K. E-mail: yuanwei.liu@qmul.ac.uk.}
	
	}

\IEEEtitleabstractindextext{%
	\begin{abstract}
	\label{abstract}
	Due to flexibility and low-cost, unmanned aerial vehicles (UAVs) are increasingly crucial for enhancing coverage and functionality of wireless networks. However, incorporating UAVs into next-generation wireless communication systems poses significant challenges, particularly in sustaining high-rate and long-range secure communications against eavesdropping attacks. In this work, we consider a UAV swarm-enabled secure surveillance network system, where a UAV swarm forms a virtual antenna array to transmit sensitive surveillance data to a remote base station (RBS) via collaborative beamforming (CB) so as to resist mobile eavesdroppers. Specifically, we formulate an aerial secure communication and energy efficiency multi-objective optimization problem (ASCEE-MOP) to maximize the secrecy rate of the system and to minimize the flight energy consumption of the UAV swarm. To address the non-convex, NP-hard and dynamic ASCEE-MOP, we propose a generative diffusion model-enabled twin delayed deep deterministic policy gradient (GDMTD3) method. Specifically, GDMTD3 leverages an innovative application of diffusion models to determine optimal excitation current weights and position decisions of UAVs. The diffusion models can better capture the complex dynamics and the trade-off of the ASCEE-MOP, thereby yielding promising solutions. Simulation results highlight the superior performance of the proposed approach compared with traditional deployment strategies and some other deep reinforcement learning (DRL) benchmarks. Moreover, performance analysis under various parameter settings of GDMTD3 and different numbers of UAVs verifies the robustness of the proposed approach.
	\end{abstract}
	
}


\maketitle

\section{Introduction}
\label{Section:Introduction}

\par \IEEEPARstart{U}{nmanned} aerial vehicles (UAVs), noted for their flexibility and low-cost, have become increasingly pivotal in various sectors, including military surveillance \cite{Samad2007}, environmental monitoring \cite{Liu2022}, and emergency response \cite{Coutinho2022}, etc. With the widespread deployment of the sixth generation (6G) wireless networks, UAVs are foreseen to play a crucial role in wireless networks as well as key enablers of innovative wireless applications \cite{Li2019a}. For instance, UAVs can serve as the mobile aerial base stations \cite{Wang2019} to support temporary and instant network coverage, which is especially valuable when the ground infrastructure is disrupted or the network capacity is insufficient to meet the demands. Moreover, UAVs can function as the aerial relays \cite{Takahashi2018} for connecting the ground users to the distant base stations and extending the coverage, particularly in rural and remote areas. Furthermore, UAVs can also access the wireless network by acting as the mobile users \cite{Zeng2019a}, enabling them to obtain real-time data and support various applications such as precision agriculture, aerial goods delivery, and environmental monitoring.

\par Although the UAVs offer significant advantages in enhancing the coverage and functionality of wireless networks, integrating them into the next-generation wireless communication and network systems also raises some crucial challenges. Specifically, maintaining high-rate and long-range communications simultaneously with a single UAV can be difficult due to the limited onboard power and potential interference \cite{Shakeri2019}. Moreover, the broadcast nature of wireless channels makes sensitive information vulnerable to eavesdropping attacks, and this vulnerability is further exacerbated in UAV-involved communications due to the high line-of-sight (LoS) probability of links \cite{Ye2024}. Although the traditional high-layer encryption and decryption techniques aim to protect data confidentiality, the advancing computing capabilities of eavesdroppers demand increasingly sophisticated algorithms, resulting in the higher computational overhead and intricate key management, which are unfeasible for UAV-involved communication systems \cite{Liu2024}.

\par Collaborative beamforming (CB) has arisen as a potential solution to the above challenges \cite{Li2021}, \cite{Zhang2024a}. Specifically, multiple UAVs can work cooperatively to construct a UAV-enabled virtual antenna array (UVAA), thereby enhancing the signal strength and directivity, which not only extends the communication range but also improves the overall secrecy rate by effectively concentrating the radiated energy in the desired direction. However, there exists a fundamental trade-off between the secure communication performance and energy consumption in the UVAA system design. In particular, to achieve an optimal beam pattern and maximize the secure transmission rate, all participating UAVs need to relocate to more suitable positions and readjust their excitation current weights, causing the increasing of the energy. Moreover, the UAVs of UVAA need to continuously adjust their positions if mobile eavesdroppers exist, which further results in additional flight energy consumption. Thus, the UVAA system must be carefully designed to balance the objectives of improving the secrecy rate of the system and reducing the flight energy consumption of the UAV swarm.

\par Traditional optimization methods, such as convex optimization \cite{Mozaffari2019} and evolutionary strategies \cite{Zhang2024a}, have been employed to deal with the optimization problems of UVAA. However, these methods may be impractical in dynamic environments due to the mobility of eavesdroppers and time-varying channel characteristics. Deep reinforcement learning (DRL) presents a compelling alternative, offering the capability to adapt to the changing conditions. It can learn optimal strategies through interactions with the environment, eliminating the need for prior knowledge and achieving near-optimal performance. Thus, DRL has been demonstrated to have great potential in wireless network optimizations \cite{Luong2019}. Nevertheless, standard DRL techniques may encounter challenges in representing the complex and high-dimensional action space required for the joint optimization of excitation current weights and positions of UAVs in UVAA. Specifically, traditional DRL methods typically use stacked fully-connected layers in the actor network, which may struggle to capture deeper data features \cite{Wang2022a}. As a result, these algorithms usually exhibit high variance, leading to a learned policy distribution that deviates from the true data distribution.

\par Recent developments in generative artificial intelligence, notably in generative diffusion models, have advanced the effective representation of complex data distributions \cite{Cao2024}. Consequently, in this study, we delve into the combination of DRL and generative diffusion models to tackle the multi-objective optimization problem in UVAA system, aimed at countering the presence of mobile eavesdroppers. The main contributions of this paper are summarized as follows:
\begin{itemize}
	\item \textbf{UAV Swarm-enabled Secure Surveillance Network System:} We propose a novel UAV swarm-enabled secure surveillance network system under the threat of mobile eavesdroppers. In this system, a UAV swarm performs CB to enhance the signal strength and directivity, thereby ensuring the secure communications between the UAV swarm and the remote base station (RBS). To the best of our knowledge, this is the first work that focuses on mobile eavesdroppers in the context of UAV-enabled CB secure communications, which is directly applicable real-world scenarios.
 
	\item \textbf{Multi-objective Optimization Problem Formulation:} We formulate an aerial secure communication and energy efficiency multi-objective optimization problem (ASCEE-MOP), with the objective of maximizing the secrecy rate between UAV swarm and RBS while minimizing the flight energy consumption of the UAV swarm by jointly optimizing the excitation current weights and positions of UAVs. Moreover, we show that the formulated ASCEE-MOP is a non-convex, NP-hard and dynamic optimization problem involving the complex trade-off, rendering it challenging to solve using traditional convex optimization techniques and evolutionary methods.
 
	\item \textbf{Generative Diffusion Model-enabled DRL Approach Design:} To deal with the non-convexity and dynamic nature of the formulated ASCEE-MOP, we re-formulate it as a Markov decision process, and address it by the DRL framework. Specifically, we propose a generative diffusion model-enabled twin delayed deep deterministic policy gradient (GDMTD3) method, which integrates the generative diffusion models within twin delayed deep deterministic policy gradient (TD3) algorithm. By utilizing the generation and inference capabilities of diffusion model, the proposed GDMTD3 can capture the complex probabilistic distribution more effectively in the high-dimensional action spaces.
 
	\item \textbf{Simulation Validation:} Simulation results are provided to demonstrate the effectiveness and robustness of the proposed approach. Specifically, compared with four deployment policies and five DRL benchmarks, the proposed approach exhibits superior performance. To further verify to the robustness, we conduct the performance analysis of the proposed GDMTD3 under various parameter settings and varying numbers of UAVs. 
\end{itemize}

\par The remainder of this paper is structured as follows. An overview of related work is provided in Section \ref{Section:Related Work}. Section \ref{Section:System Model} outlines the system model. Next, the optimization problem is formulated and analyzed in Section \ref{Section:Problem Formulation and Analysis}. Section \ref{Section:The Proposed GDMTD3} details the GDMTD3 for addressing the formulated optimization problem. Simulation results are listed and discussed in Section \ref{Section:Simulation Results}, and the conclusion of the paper is presented in Section \ref{Section:Conclusion}.

\begin{table*}[ht]
	\centering
	\renewcommand{\arraystretch}{1.3}
	\caption {Major Notions}
	\label{Tab:Major Notions}
	\begin{tabular}[c]{p{0.1cm}|cl|cl}
		\toprule[1.5pt]
		& {\textbf{Symbols}} & {\textbf{Definition}} & {\textbf{Symbols}} & {\textbf{Definition}} \\
		\toprule[1pt]
		\multirow{18}{*}{\rotatebox{90}{\textbf{System Model}}}
		& $\mathcal{K}$ & Set of UAV indexes, $|\mathcal{K}| = K$  & $\boldsymbol{w}_{B}$ & Coordinate of BS \\
		& $N$ & Total number of time slots  & $\boldsymbol{q}_{k}^{U}$ & Coordinate of UAV $k$ \\
		& $I_{k}^{U}$ & Excitation current weight of UAV $k$ & $\boldsymbol{q}_{c}$ & Coordinate of UVAA center \\
		& $\theta,\varphi$ & Elevation and azimuth angles  & $\boldsymbol{q}_{E}$ & Coordinate of mobile eavesdropper \\
		& $AF$ & Array factor of UVAA  & $\Psi_{k}$ & Initial phase of UAV $k$ \\
		& $c_{0},c_{1}$ & Two constants depending on wireless environment  & $c_{p}$ & Phase constant \\
		& $d_{c,\mathcal{S}}, d_{c,E}$ & Distances between UVAA and BS/eavesdropper  & $\lambda$ & Wavelength \\
		& $P_{c,\mathcal{S}}^{\text{LoS}}, P_{c,E}^{\text{LoS}}$ & LoS link probability between UVAA and BS/eavesdropper  & $c, f_{c}$ & Light speed and Carrier frequency \\
		& $\overline{L}_{c,\mathcal{S}}$ & Average pass loss between UVAA and BS  & $\xi$ & Elevation between UVAA and BS \\
		& $g_{c,\mathcal{S}}, g_{c,\mathcal{E}}$ & Channel gain between UVAA and BS/eavesdropper  & $\mu_{1},\mu_{2}$ & Excessive path loss for LoS and NLoS links \\
		& $G_{U,\mathcal{S}}, G_{U,E}$ & Antenna gain of UVAA towards BS/eavesdropper  & $\alpha$ & Path loss exponent \\
		& $R_{U,\mathcal{S}}, R_{U,E}$ & Transmission rate from UVAA to BS/eavesdropper  & $B$ & Transmission bandwidth \\
		& $\sigma^{2}$ & Noise power of A2G channel & $R_{SE}$ & Achievable secrecy rate of A2G link \\
		& $v_{k}^{x}, v_{k}^{y}, v_{k}^{z}$ & $x/y/z$-axis component speed of the UAV $k$  & $\rho$ & Density of air \\
		& $W$ & Weight of UAV  & $A$ & Total area of UAV rotor disks \\
		& $v_{0}$ & Mean rotor induced velocity for hovering  & $d_{0}$ & Fuselage drag ratio \\
		& $s$ & Rotor solidity & $P_\text{level}^{k}$ & Induced power of UAV $k$ for level flight \\
		& $P_\text{vertical}^{k}$ & Power of UAV $k$ for vertical flight & $E$ & Energy consumption of UAV swarm\\
		\midrule[1.5pt]
		\multirow{6}{*}{\rotatebox{90}{\textbf{Algorithm}}} & $\mathcal{S}, \boldsymbol{s}$ &  State space and state vector of environment & $\mathcal{A}, \boldsymbol{a}$ & Action space and action vector of agent \\
		& $\mathcal{P}$ & State transition probability of environment & $\mathcal{R}, r$ & Reward space and reward \\
		& $\gamma$ & Discount factor & $d$ & Frequency of policy update \\
		& $\boldsymbol{\theta_{Q_{i}}}, \boldsymbol{\theta_{Q_{i}}^{\prime}}$ & Parameters of the $i$th critic network and target critic network & $\boldsymbol{Q}(\boldsymbol{s}, \boldsymbol{a})$ & State-action value function\\
		& $\boldsymbol{\theta_{d}}, \boldsymbol{\theta_{d}^{\prime}}$ & Parameters of actor network and target actor network & $\boldsymbol{\kappa_{\theta_{d}}}(\bm{x}_t, t, \boldsymbol{g})$ & Mean function of diffusion reverse process  \\
		& $\boldsymbol{x_{t}}$ & Noisy sample at the $t$th denoising step & $\tilde{\beta}_t$ & Predetermined variance factor \\
		\bottomrule[1.5pt]
	\end{tabular}
\end{table*}

\par \textit{Notations}: We use plain symbols to stand for scalars (e.g., $a, b$), bold symbols for vectors or functions (e.g., $\boldsymbol{a},  \boldsymbol{b}$), and calligraphic symbols for sets (e.g., $\mathcal{A}, \mathcal{B}$).  $\Vert \cdot \Vert$ represents Euclidean norm, and $\left\{\cdot\right\}^{+}$ refers to $\max\{0,\cdot\}$. Accordingly, Table~\ref{Tab:Major Notions} outlines the major notions adopted in the following sections.

\section{Related Work}
\label{Section:Related Work}

\par In this section, we discuss related works on UAV-enabled secure communications, optimization objectives in aerial secure communications, and optimization methods for aerial secure communications.

\subsection{UAV-enabled Aerial Secure Communications}
\label{SubSection:UAV-enabled Aerial Secure Communications}

\par A number of prior works have concentrated on utilizing UAVs to enhance the security performance of wireless communications. In terms of the number of UAVs, the existing works can primarily be categorized into the single UAV-enabled secure communications and multiple UAVs-enabled secure communications.

\par For the single UAV-enabled secure communications, Zhang \textit{et al}. \cite{Zhang2019} investigated the security of both UAV-to-ground and ground-to-UAV communications to mitigate the risk posed by an stationary eavesdropper. Cheng \textit{et al}. \cite{Cheng2019} introduced a secure scheme to maximize the secrecy rate of the UAV-enabled wireless relay networks with caching, where a UAV is employed to relay the data from the base station to the users, leveraging its mobility. In \cite{Zhou2020}, the authors considered a secure UAV mobile edge computing system, where a legitimate UAV assists in processing large computing tasks offloaded from multiple ground users in the presence of multiple eavesdropping UAVs. Moreover, Sun \textit{et al}. \cite{Sun2020} explored UAV-enabled downlink mmWave simultaneous wireless information and power transfer (SWIPT) networks, involving two types of authorized users with different communication needs and multiple passive eavesdroppers modeled by independent homogeneous Poisson point processes. In \cite{Li2019}, the authors studied a UAV-enabled mobile jamming strategy to enhance the secrecy rate of ground wiretap channels.

\par For multiple UAVs-enabled secure communications, Cai \textit{et al}. \cite{Cai2020} explored a joint optimization strategy for the trajectory and resource allocation of the UAV communication systems. In their approach, one UAV acts as an information transmitter while another one serves as an assisting jammer to enhance the energy efficiency and security. In \cite{Gao2024}, the authors presented a dynamic role-switching strategy, where the UAVs act as data collectors or jammers based on their locations to serve multiple ground users. Hanna \textit{et al}. \cite{Hanna2023} achieved the reliable beamforming by considering estimation errors and employing a Kalman filter for frequency tracking, with validation through simulations and experiments on software-defined radios and UAVs.

\par However, these aforementioned works focus on non-remote communication settings due to the limited energy of UAVs. Moreover, they primarily consider secure communication scenarios involving static eavesdroppers.

\subsection{Optimization Objectives in Aerial Secure Communications}
\label{SubSection:Optimization Objectives in Aerial Secure Communications}

\par Optimization objectives have a significant role in enhancing the performance and security of UAV-enabled secure communications. Previous research has predominantly concentrated on two aspects that are the secrecy rate and flight energy consumption of UAVs.

\par The secrecy rate is a key metric for measuring communication security, representing the maximum achievable confidential transmission rate in the existence of potential eavesdroppers. Several studies are dedicated to maximizing the secrecy rate in UAV-enabled secure communication systems. For example, in \cite{Mamaghani2023}, the authors studied a secure short-packet communication system by using a UAV as the mobile relay. Specifically, they jointly optimized the coding blocklengths, transmit powers, and UAV trajectory to enhance the secrecy throughput. Fan \textit{et al}. \cite{Fan2020} proposed an iterative algorithm to optimize the UAV trajectory, transmit power, and user scheduling for achieving secure communications, addressing eavesdropper position estimation errors and ensuring user service fairness. In \cite{Gao2021}, the authors investigated an iterative suboptimal algorithm to maximize the worst average secrecy rate in the UAV-enabled networks by optimizing the UAV trajectory, transmit power, and user scheduling while considering energy constraints and security threats from external and internal eavesdroppers. 

\par Several studies take into account the flight energy consumption of UAVs due to the limited battery capacity. For example, Gao \textit{et al}. \cite{Gao2021a} aimed to minimize the energy consumption of a fixed-wing UAV under security constraints, where they jointly optimized user scheduling and UAV trajectory in a scenario with multiple colluding eavesdroppers. In \cite{Mao2023}, the authors formulated an energy consumption minimization problem subject to constraints such as users service quality and information security requirements by jointly optimizing the offloading time, CPU frequency, artificial noise, beamforming vectors, and trajectory of UAV, along with the offloading time, CPU frequency, and transmit power of each user.

\par However, there exists a clear trade-off between maximizing the secrecy rate and minimizing flight energy consumption, especially in UAV-enabled CB communication systems. In such systems, each individual in the UAV swarm must continuously adjust its position to enhance the directivity of UVAA. Dong \textit{et al}. \cite{Dong2021} considered a UVAA-enabled relay system, where they focused on maximizing achievable secrecy rate of downlink by jointly optimizing the beamforming vector of
UVAA and bandwidth allocation. Although this process improves the security performance compared to a single UAV-enabled secure communications, it also results in the increased flight energy consumption. To deal with this trade-off, we formulate a multi-objective optimization problem that seeks to maximize the secrecy rate of system and minimize the flight energy consumption of the UAV swarm by jointly optimizing the excitation current weights and positions of UAVs.

\subsection{Optimization Methods for Aerial Secure Communications}
\label{SubSection:Optimization Methods for Aerial Secure Communications}

\par To address the optimization problems for the UAV-enabled secure communication systems, researchers are devoted to effective algorithm design by employing methodologies such as convex optimization, swarm intelligent and DRL methods. For example, Zhou \textit{et al}. \cite{Zhou2019} utilized the successive convex approximation to solve the joint optimization problem of the transmit powers and trajectories of UAV jammer and aerial base station. Furthermore, Li \textit{et al}. \cite{Li2021} proposed an improved multi-objective dragonfly algorithm with chaotic solution initialization and  (IMODACH) to deal with the trade-off among the secrecy rate and maximum sidelobe level and energy consumption in UAV-enabled secure communications. Moreover, Xiao \textit{et al}. \cite{Xiao2022a} developed a hierarchical DRL algorithm to enhance the anti-eavesdropping performance, with regard to the outage probability, intercept probability, energy consumption and latency. Moreover, in \cite{Dong2023}, the authors utilized a modified proximal policy optimization method to minimize the secrecy outage duration and the weighted sum of flight period by jointly optimizing the UAV trajectory, the user scheduling and the beamforming vector. 

\par However, both convex optimization and swarm intelligence methods have certain limitations in their applicability to dynamic environments. Therefore, we explore DRL method to deal with the formulated optimization problem. Despite the potential advantages of many DRL-based methods in dynamic environments, they still face limitations in handling the complexities and uncertainties of dynamic environments. To address this issue, our work integrates the generative diffusion model with DRL, thereby improving the ability of the algorithm to model more complex probabilistic distribution in high-dimensional action spaces.

\section{System Model}
\label{Section:System Model}

\par In this section, we first present a comprehensive system description. Subsequently, we delve into the details of the considered models, including the array factor, channel gain, secrecy rate, and UAV energy consumption models.

\subsection{System Description}
\label{SubSection:System Description}

\begin{figure}[t]
	\centering	\includegraphics[width=\linewidth,scale=1.00]{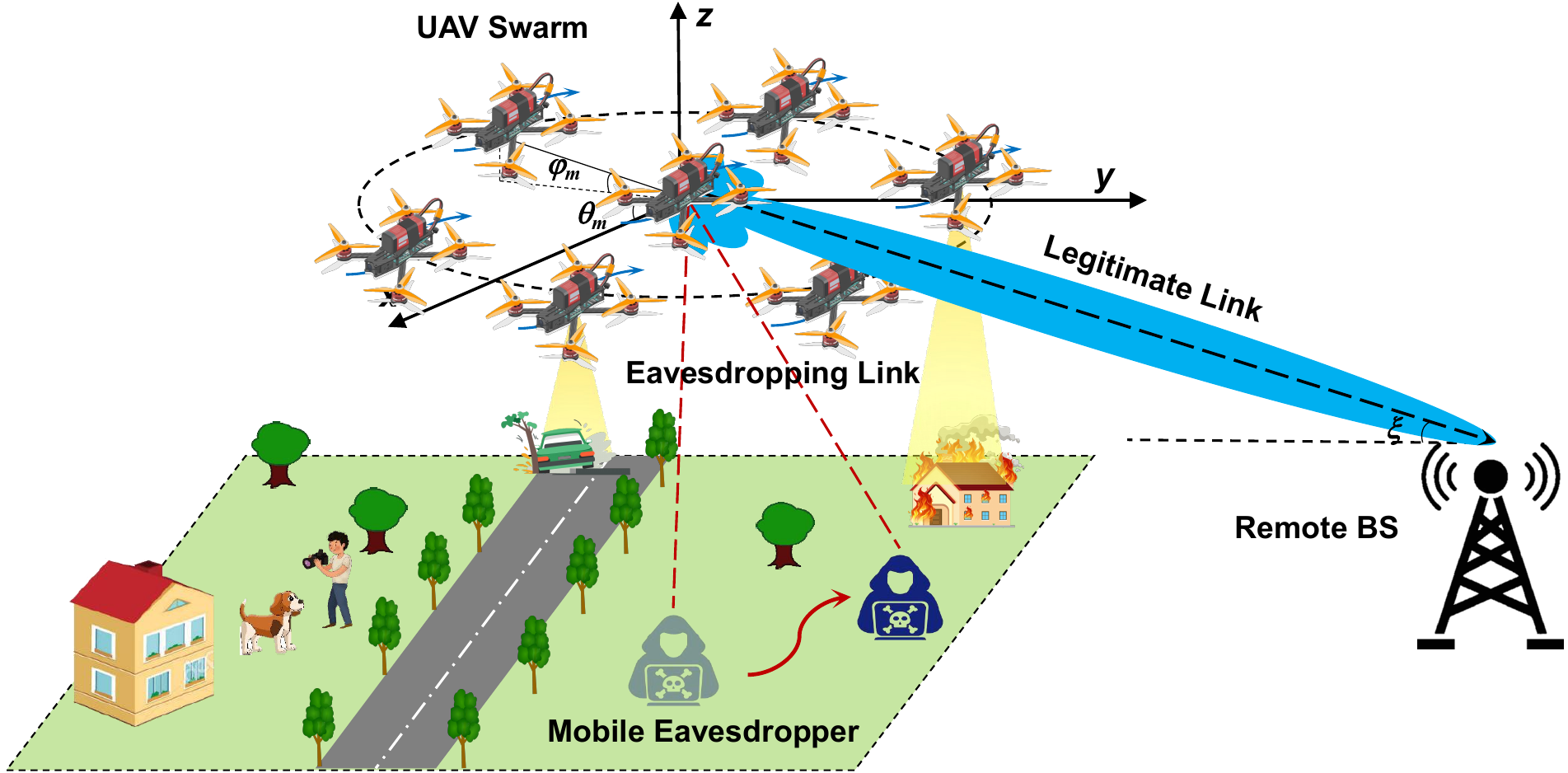}
	\caption{A UAV swarm-enabled secure surveillance network system, where a UAV swarm is deployed for surveillance tasks, transmitting sensitive data to a RBS. The security of system is challenged by a mobile eavesdropper, depicted by red dashed lines, attempting to intercept the data via wiretap links over various time slots.}
	\label{Figure:System Model}
\end{figure}

\par As shown in Fig.~\ref{Figure:System Model}, we consider a UAV swarm-enabled secure surveillance network system, which consists of $K$ UAVs denoted by $\mathcal{K} \triangleq \{1, 2, \cdots, K\}$ and one RBS denoted by $\mathcal{S}$. Specifically, the UAVs have collected some sensitive surveillance data and need to transmit the data back to the RBS $\mathcal{S}$ by wireless links over a given time period $T$. For ease of exposition, the total time $T$ is further divided into $N$ time slots with equal duration $\delta_{t}$, i.e., $T \triangleq N\delta_{t}$. However, due to blockage of obstacles and signal attenuation for long distance communication, a single power-constrained UAV is not able to send data to RBS $\mathcal{S}$ directly. Moreover, there exists a mobile eavesdropper on the ground trying to intercept the sensitive information. To enhance the transmission efficiency and resist eavesdropping attacks from the mobile eavesdropper, these UAVs will form a UVAA to perform CB and transmit data back to RBS $\mathcal{S}$ on the air-to-ground (A2G) link.

\par Mathematically, all entities are defined within a three-dimensional Cartesian coordinate system. Specifically, the RBS $\mathcal{S}$ is situated at a fixed point denoted by $\boldsymbol{w}_{B} = \left(x_{\mathcal{S}}, y_{\mathcal{S}}, H_{\mathcal{S}}\right)$. Moreover, it is worth noting that the position change of UAVs and eavesdropper within a time slot can be negligible since the duration $\delta_{t}$ is chosen to be sufficiently small. Thus, the $3$D coordinates of UAV $k$ and mobile eavesdropper at time slot $n$ are denoted by $\boldsymbol{q}_{k}^{U}[n] = \left(x_{k}^{U}[n],  y_{k}^{U}[n], z_{k}^{U}[n]\right)$ and $\boldsymbol{q}_{E}[n] = \left(x_{E}[n], y_{E}[n], 0\right)$, respectively.

\subsection{Array Factor Model}
\label{SubSection:Array Factor Model}

\par The virtual antenna array formed by UAV swarm can significantly improve the antenna directivity by optimizing its beam pattern. Specifically, at time slot $n$, the excitation current weight of UAV $k$ is denoted as $I_{k}^{U}[n]$, the coordinate of UVAA center $\boldsymbol{q}_{c}[n] = \left(x_{c}^{U}[n], y_{c}^{U}[n], z_{c}^{U}[n]\right)$, and the component distances in the $x$-axis, $y$-axis and $z$-axis between UAV $k$ and UVAA center are represented by $d_{c, k}^{x}[n], d_{c, k}^{y}[n]$ and $d_{c, k}^{z}[n]$, respectively. According to electromagnetic wave superposition principle, the array factor (AF) of UVAA at time slot $n$ can be described as follows \cite{Li2024}:
\begin{equation}
    \begin{aligned}
        AF&\left(\theta, \varphi~|~\theta_{\mathcal{S}}[n], \varphi_{\mathcal{S}}[n]\right)=\sum_{k=1}^{K}\Big(I^{U}_{k}[n]e^{\Psi_{k}\left(\theta_{\mathcal{S}}[n], \varphi_{\mathcal{S}}[n]\right)}\\&\cdot e^{j\left[c_{p}\left(d_{c, k}^{x}[n]\operatorname{sin}\theta\operatorname{cos}\varphi+d_{c,k}^{y}[n]\operatorname{sin}\theta\operatorname{sin}\varphi+d_{c,k}^{z}[n]\operatorname{cos}\theta\right)\right]}\Big),
    \end{aligned}
    \label{Equation:Array Factor of UVAA}
\end{equation}

\noindent where $\lambda$ is the wavelength, and $c_{p} = 2\pi/\lambda$ is the phase constant. Moreover, $\theta \in [0, \pi]$ and $\varphi \in [-\pi, \pi]$ are the elevation and azimuth angles, respectively. In addition, the direction of RBS $\mathcal{S}$ with respect to UVAA $\boldsymbol{q}_{c}[n]$ is denoted as $(\theta_{\mathcal{S}}[n], \varphi_{\mathcal{S}}[n])$ at time slot $n$, and $\Psi_{k}\left(\theta_{\mathcal{S}}[n], \varphi_{\mathcal{S}}[n]\right)$ is the initial phase of UAV $k$ in UVAA at time slot $n$. 

\par In this work, we adopt an open-loop phase synchronization scheme \cite{Ochiai2005}, which can be easily implemented through UAV swarm intra-cluster communication protocols \cite{Feng2013}. For this case, the initial phase synchronization is accomplished by offsetting the distance between the UAV and UVAA center. As a result, the initial phase of UAV $k$ in UVAA can be calculated as follows:
\begin{equation}
    \begin{aligned}
        \Psi_{k}\left(\theta_{\mathcal{S}}[n], \varphi_{\mathcal{S}}[n]\right) = -c_{p}\Big(&d_{c, k}^{x}[n]\operatorname{sin}\theta_{\mathcal{S}}[n]\operatorname{cos}\varphi_{\mathcal{S}}[n]\\&+d_{c,k}^{y}[n]\operatorname{sin}\theta_{\mathcal{S}}[n]\operatorname{sin}\varphi_{\mathcal{S}}[n]\\&+d_{c,k}^{z}[n]\operatorname{cos}\theta_{\mathcal{S}}[n]\Big).		
    \end{aligned}
    \label{Equation:Initial phase of UVAA}
\end{equation}

\subsection{Channel Gain Model}
\label{SubSection:Channel Gain Model}

\par To precisely model the A2G wireless communications, we utilize the elevation angle-dependent probabilistic LoS model \cite{AlHourani2014} to  characterize the A2G communication between UVAA and RBS $\mathcal{S}$. Specifically, the LoS link probability between UVAA and RBS $\mathcal{S}$ at time slot $n$ can be given by
\begin{equation}
	P_{c,\mathcal{S}}^{\text{LoS}}[n] = \frac{1}{1+c_{0}\operatorname{exp}\left(-c_{1}\left(\xi[n]-c_{0}\right)\right)},
	\label{Equation:The probability of LoS between UVAA and BS}
\end{equation}

\noindent where $c_{0}$ and $c_{1}$ are two constants depending on the carrier frequency and environment. As depicted in Fig.~\ref{Figure:System Model}, $\xi[n]$ is the elevation between UVAA center and RBS $\mathcal{S}$ at time slot $n$ and can be calculated by $ \frac{180}{\pi}\operatorname{arcsin}\left(\frac{z_{c}^{U}[n]-H_{\mathcal{S}}}{d_{c,\mathcal{S}}[n]}\right)$, wherein $d_{c,\mathcal{S}}[n] = \sqrt{\Vert \boldsymbol{q}_{c}[n] - \boldsymbol{w}_{B}\Vert^{2}}$ is the distance between UVAA center and RBS $\mathcal{S}$ at time slot $n$. Accordingly, the NLoS link probability at time slot $n$ can be expressed as $P_{c,\mathcal{S}}^{\text{NLoS}}[n]=1-P_{c,\mathcal{S}}^{\text{LoS}}[n]$.

\par Thus, the path loss for LoS and NLoS links between UVAA and RBS $\mathcal{S}$ at time slot $n$ can be given by \cite{Nobar2022}
\begin{equation}
	L_{c,\mathcal{S}}[n]=\left\{
	\begin{aligned}
		\mu_{1}\left(\frac{4\pi f_{c} d_{c,\mathcal{S}}[n]}{c}\right)^\alpha, &\quad \text{LoS~link}\\
		\mu_{2}\left(\frac{4\pi f_{c} d_{c,\mathcal{S}}[n]}{c}\right)^\alpha, &\quad \text{NLoS~link}\\
	\end{aligned}
	\right.,
	\label{Equation:LoS and NLoS path loss between UVAA and BS}
\end{equation}

\noindent where $\mu_{1}$ and $\mu_{2}$ $\left(\mu_{2} > \mu_{1} > 1\right)$ represent the excessive path loss for LoS and NLoS links, respectively. Moreover, $c$ is the light speed, $\alpha$ is the path loss exponent, and $f_{c}$ is the carrier frequency. 

\par Typically, considering both LoS and NLoS links, the average pass loss between UVAA and RBS $\mathcal{S}$ at time slot $n$ can be express as follows:
\begin{equation}
	\overline{L}_{c,\mathcal{S}}[n] = \left[P_{c,\mathcal{S}}^{\text{LoS}}[n]\mu_{1} + P_{c,\mathcal{S}}^{\text{NLoS}}[n]\mu_{2}\right]\left(K_{o}d_{c,\mathcal{S}}[n]\right)^{\alpha},
	\label{Equation:The average path loss between UVAA and BS}
\end{equation}

\noindent where $K_{o} = \frac{4\pi f_{c}}{c}$ represents the free-space path loss factor. Furthermore, the channel gain between UVAA center and RBS $\mathcal{S}$ at time slot $n$ can be calculated as $g_{c,\mathcal{S}}[n]=\frac{1}{\overline{L}_{c,\mathcal{S}}[n]}$.

\par Similarly, the channel gain between UVAA and mobile eavesdropper at time slot $n$ is described as follows:
\begin{equation}
	{g}_{c,E}[n] =\frac{1}{ \left[P_{c,E}^{\text{LoS}}[n]\mu_{1} + P_{c,E}^{\text{NLoS}}[n]\mu_{2}\right]\left(K_{o}d_{c,E}[n]\right)^{\alpha}},
	\label{Equation:The average path loss between UVAA and Eavesdropepr}
\end{equation}

\noindent where $P_{c,E}^{\text{LoS}}[n]$ and $P_{c,E}^{\text{NLoS}}[n]$ represent the probabilities of LoS and NLoS links between UVAA and mobile eavesdropper at time slot $n$, respectively. Moreover, $d_{c,E}[n]$ is the distance between UVAA center and mobile eavesdropper at time slot $n$, which can be calculated by $d_{c,E}[n] = \sqrt{\Vert \boldsymbol{q}_{c}[n] - \boldsymbol{q}_{E}[n]\Vert^{2}}$.

\subsection{Secrecy Rate Model}
\label{SubSection:Secrecy Rate Model}

\par By exploiting the previously mentioned array factor and channel model, the transmission rate from UVAA to RBS at time slot $n$ can be expressed as follows:
\begin{equation}
	R_{U,\mathcal{S}}[n] = \operatorname{log}_{2}\left(1 + \frac{P_{U}[n]g_{c,\mathcal{S}}[n]G_{U,\mathcal{S}}(\theta_{\mathcal{S}}[n],\varphi_{\mathcal{S}}[n])}{\sigma^2}\right),
	\label{Equation:The transmission rate between UVAA and BS}
\end{equation}

\noindent where $P_{U}[n]$ represents the transmit power of UVAA, and $\sigma^2$ is the noise power of the A2G channel. Moreover, $G_{U,\mathcal{S}}(\theta_{\mathcal{S}}[n],\varphi_{\mathcal{S}}[n])$ is the antenna gain\footnote{In this work, we assume that the magnitude of
 the far-field beam pattern of each UAV element is $0$ dB since each UAV is equipped with a single isotropic antenna under the same power constraints. Moreover, the antenna efficiency is approximated as to be $1$.} of UVAA towards RBS $\mathcal{S}$ at time slot $n$, which can be defined as follows:
\begin{equation}
    \begin{aligned}
        G_{U,\mathcal{S}}&(\theta_{\mathcal{S}}[n],\varphi_{\mathcal{S}}[n])=\\&\frac{4 \pi\left|AF\left(\theta_{\mathcal{S}}[n], \varphi_{\mathcal{S}}[n]|\theta_{\mathcal{S}}[n], \varphi_{\mathcal{S}}[n]\right)\right|^2}{\int_0^{2 \pi} \int_0^\pi|AF(\theta, \varphi|\theta_{\mathcal{S}}[n], \varphi_{\mathcal{S}}[n])|^2 \sin \theta \mathrm{d} \theta \mathrm{d} \varphi}.
    \end{aligned}
    \label{Equation:Antenna gain towards BS}
\end{equation}

\par Similarly, the antenna gain of UVAA towards the mobile eavesdropper at time slot $n$ can be written as follows:
\begin{equation}
    \begin{aligned}
        G_{U,E}&(\theta_{E}[n],\varphi_{E}[n])=\\&\frac{4 \pi\left|AF\left(\theta_{E}[n], \varphi_{E}[n]|\theta_{0}[n], \varphi_{0}[n]\right)\right|^2}{\int_0^{2 \pi} \int_0^\pi|AF(\theta, \varphi|\theta_{0}[n], \varphi_{0}[n])|^2 \sin \theta \mathrm{d} \theta \mathrm{d} \varphi},
    \end{aligned}
    \label{Equation:Antenna gain towards Eavesdropper}
\end{equation}

\noindent where $\left(\theta_{E}[n], \varphi_{E}[n]\right)$ is the direction of the mobile eavesdropper with respect to the UVAA center at time slot $n$. Accordingly, the transmission rate from UVAA to the mobile eavesdropper can be expressed as follows:
\begin{equation}
	R_{U,E}[n] = \operatorname{log}_{2}\left(1 + \frac{P_{U}[n]g_{c,E}[n]G_{U,E}(\theta_{0}[n],\varphi_{0}[n])}{\sigma^2}\right).
	\label{Equation:The transmission rate between UVAA and Eavesdropper}
\end{equation}

\par Furthermore, the achievable secrecy rate of A2G wireless link at time slot $n$ is given by
\begin{equation}
	\begin{aligned}
		R_{SE}[n] &= \left\{R_{U,\mathcal{S}}[n] - R_{U,E}[n]\right\}^{+},\\
	\end{aligned}
	\label{Equation:The secure rate}
\end{equation}

\noindent where $\left\{x\right\}^{+}$ is defined as $\operatorname{max}\{x,0\}$.

\subsection{UAV Energy Consumption Model}
\label{SubSection:UAV Energy Consumption Model}

\par According to the aircraft dynamics of rotary-wing UAVs, the power consumption can be expressed as the sum of the power for level flight and the power for vertical flight \cite{Meng2022}. Specifically, the power of UAV $k$ for level flight at time slot $n$ can be calculated as follows:
\begin{equation}
	\begin{aligned}
		P_\text{level}^{k}[n]= &P_{i} \sqrt{\sqrt{1+\frac{\Vert v_{k}^{x}[n], v_{k}^{y}[n]\Vert^{4}}{4v_{0}^{4}}} - \frac{\Vert v_{k}^{x}[n], v_{k}^{y}[n]\Vert^{2}}{2v_{0}^{2}}}\\
		& + P_{0}\left(1 + \frac{3\Vert v_{k}^{x}[n], v_{k}^{y}[n]\Vert^{2}}{u_{tip}^{2}}\right)\\
		& + \frac{1}{2}d_{0}\rho s A \Vert v_{k}^{x}[n], v_{k}^{y}[n]\Vert^{3},
	\end{aligned}
	\label{Equation:UAV power for level flight}
\end{equation}

\noindent where $v_{k}^{x}$ and $v_{k}^{y}$ are the $x$-axis component speed and $y$-axis component speed of UAV $k$ at time slot $n$, respectively. $v_{0}$ is the mean rotor induced velocity for hovering, $U_{tip}$ is the tip speed of the rotor blade, $d_{0}$ is the fuselage drag ratio, $\rho$ is the density of air, $s$ is the rotor solidity and $A$ is the rotor disk area. Moreover, $P_{i}$ and $P_{0}$ denote the induced power and the blade profile power in hovering status, which can be calculated as follows \cite{Zeng2019}:
\begin{equation}
	P_{i} = (1+M)\frac{W^{3/2}}{\sqrt{2\rho A}}, P_{0} = \frac{\kappa}{8}\rho s A \Omega^{3}\Lambda^{3},
	\label{Equation:UAV induced and blade power}
\end{equation}

\noindent where $\Omega$ is the blade angular velocity, $M$ is the incremental correction factor to induced power, $\Lambda$ is the rotor radius, and $\kappa$ is the profile drag coefficient. Moreover, $W=mg$ is the weight of UAV, wherein $g$ is gravitational acceleration and $m$ is the mass of UAV.

\par In addition, the power of UAV $k$ for vertical flight at time slot $n$ can be modeled as follows:
\begin{equation}
	P_\text{vertical}^{k}[n]= \left\{
	\begin{aligned}
		&Wv_{k}^{z}[n], && v_{k}^{z}[n] > 0\\
		&0, && v_{k}^{z}[n] \leq 0
	\end{aligned}
	\right.,
	\label{Equation:UAV power for vertical flight}
\end{equation}

\noindent where $v_{k}^{z}$ is the $z$-axis component speed of UAV $k$ at time slot $n$. Moreover, $P_\text{vertical}^{k}[n]=0$ as the UAVs operate in
auto-rotation and are unpowered during the vertical descent \cite{Meng2022}.

\par Accordingly, the flight energy consumption of UAV swarm at time slot $n$ can be modeled as follows:
\begin{equation}
	E[n] = \sum_{k=1}^{K}\delta_{t}(P_\text{level}^{k}[n] + P_\text{vertical}^{k}[n]).	
	\label{Equation:UAV power consumption}	
\end{equation}

\section{Problem Formulation and Analysis}
\label{Section:Problem Formulation and Analysis}

\par In this work, we aim to maximize the secrecy rate of the system while minimizing the flight energy consumption of the UAV swarm by determining the excitation current weights and positions of UAVs during a period of $N$ time slots. Thus, the ASCEE-MOP is formulated as follows: 
\begin{subequations}
	\label{Problem Formulation}
	\begin{align}
		\textbf{P1:} \quad& \underset{\boldsymbol{I},  \boldsymbol{q}}{\text{max}}\  (\sum_{n=1}^{N} R_{SE}[n], -\sum_{n=1}^{N}E[n]), \label{Equation:Optimization Objective}\\
		\text{s.t.}\ &0 \leq I_{k}^{U}[n] \leq 1, \forall k \in \{1,...,K \},\label{Constraint:Excitaion Current Weight}\\
		&X_{min} \leq x_{k}^{U}[n] \leq X_{max}, \forall k \in \{1,...,K \},\label{Constraint:UAV Bound X}\\
		&Y_{min} \leq y_{k}^{U}[n] \leq Y_{max}, \forall k \in \{1,...,K \},\label{Constraint:UAV Bound Y}\\
		&Z_{min} \leq z_{k}^{U}[n] \leq Z_{max}, \forall k \in \{1,...,K \},
		\label{Constraint:UAV Bound Z}\\
		&0 \leq v_{k}^{U}[n] \leq V_{max}, \forall k \in \{1,...,K \},\label{Constraint:UAV Speed}\\
		&\Vert \boldsymbol{q}_{k_{1}}[n], \boldsymbol{q}_{k_{2}}[n] \Vert\ge D_{min}^{U}, \forall k_{1}, k_{2} \in \{1,...,K\},\label{Constraint:UAV Space}
	\end{align}
\end{subequations}

\noindent where $\boldsymbol{I}[n]$ and $\boldsymbol{q}[n]$ are the excitation current weights and positions of UAVs at time slot $n$, respectively. Constraint \eqref{Constraint:Excitaion Current Weight} expresses the range constraint of the excitation current weight. Moreover, Constraints \eqref{Constraint:UAV Bound X}, \eqref{Constraint:UAV Bound Y} and \eqref{Constraint:UAV Bound Z} restrict the flight area of the UAV which may be imposed by surveillance area and government regulations. In addition, Constraint \eqref{Constraint:UAV Speed} is the speed constrain of the UAV, and Constraint \eqref{Constraint:UAV Space} is imposed to guarantee the minimum distance between two UAVs.

\par \underline{\textit{\textbf{Non-convexity:}}} The ASCEE-MOP is inherently non-convex, stemming from both its imposed safety constraints and objective function. Specifically, the safety constraint, as delineated in Constraint \eqref{Constraint:UAV Space}, necessitates a minimum separation distance between UAVs, thereby resulting in a non-convex solution space defined by regions external to spherical boundaries.

\par \underline{\textit{\textbf{NP-hard:}}} The formulated ASCEE-MOP can be proven to be NP-hard. Specifically, we assume that the optimization problem is simplified by only considering to maximize the secrecy rate of system at a given time slot with fixing the positions of UAVs. Moreover, the excitation current weights are further simplified as the discrete values, i.e., $I_{k}^{U} \in \boldsymbol{S} =\{0, 1\}$. Accordingly, the simplified problem is given as follows:

\begin{subequations}
	\label{NP-hard}
	\begin{align}
	\textbf{P2:}\quad& \underset{\boldsymbol{I}}{\text{max}}\quad R_{SE}, \\
		\text{s.t.}\ & I_{k}^{U} \in \boldsymbol{S} , \forall k \in \{1,...,K \},\\
        &\sum_{k=1}^{K}{I_{k}^{U}} \leq K, \forall k \in \{1,...,K \},
	\end{align}
\end{subequations}

\par As such, the $\textbf{P2}$ is structured as a nonlinear multi-dimensional knapsack problem, which is NP-hard \cite{Goos2020}. Therefore, the ASCEE-MOP is an NP-hard optimization problem since it is much more complex than \textbf{P2}.

\par \underline{\textit{\textbf{Trade-off:}}} Furthermore, the objective function of ASCEE-MOP seeks to concurrently maximize the secrecy rate of the system while minimizing the flight energy consumption of the UAV swarm. Specifically, it is essential for UAVs to fly to suitable positions to improve the antenna directivity of the UVAA system, thereby maximizing the total secrecy rate during task execution. However, constantly adjusting the positions of UAVs to maintain optimal antenna directivity leads to significant energy consumption. Thus, there is an inherent trade-off between maximizing the secrecy rate of the system and minimizing flight energy consumption of the UAV swarm within the formulated ASCEE-MOP, and striking the right balance between these two conflicting objectives poses a challenging task.

\par To deal with such non-convex optimization problems, most works subdivide them into several convex subproblems which can be solved by an iterative manner. However, the accuracy is impacted as a result of the decomposition. Moreover, the dynamics of environment, e.g., the changed position of mobile eavesdropper and the time-varying channel, brings some challenges. In this case, existing optimization-based methods and heuristic algorithms needs to re-run once the environment changes. Fortunately, DRL provides a feasible and efficient way for the sequential decision making and optimal control in dynamic environments. Thus, this motives us to utilize DRL-based methods to address the formulated ASCEE-MOP.

\section{The Proposed GDMTD3}
\label{Section:The Proposed GDMTD3}

\par In this section, the formulated non-convex multi-objective optimization problem is solved by the DRL-based method. Specifically, we first adopt a Markov decision process to reformulate the ASCEE-MOP, and then propose the GDMTD3 method to solve the problem.

\subsection{Markov Decision Process for ASCEE-MOP}
\label{SubSection:Markov Decision Process for ASCEE-MOP}

\par The formulated ASCEE-MOP of the UAV swarm-enabled surveillance network system can be modeled as a Markov decision process to facilitate the application of DRL. In general, a Markov decision process is represented as a tuple $<\mathcal{S},\mathcal{A}, \mathcal{P}, \mathcal{R}, \gamma>$, where $\mathcal{S}$ is the state space of environment, $\mathcal{A}$ is the action space of agent, $\mathcal{P}$ denotes the state transition probability of environment, $\mathcal{R}$ is the reward space, and $\gamma \in [0, 1]$ denotes the reward discount factor. Specifically, the UVAA is treated as a decision-making agent in the Markov decision process. With the framework of the Markov decision process, the environment state at any given time slot $n$ is signified by $\boldsymbol{s[n]}$, wherein $\boldsymbol{s}[n] \in \mathcal{S}$. Subsequently, the agent selects an action $\boldsymbol{a}[n]$ according to the policy $\boldsymbol{\pi}(\boldsymbol{s}[n])$. After that, the environment dispenses the agent a reward $r[n]$ and transitions to the next state $\boldsymbol{s}[n+1]$ based on the transition probability function $\mathcal{P}(\boldsymbol{s}[n+1]|\boldsymbol{s}[n], \boldsymbol{a}[n])$. Accordingly, the crucial elements in our model are described below in detail.

\subsubsection{\textbf{State Space}}
\label{SubSubSection: State Space}

\par The state of the system at time slot $n$ can be defined by $\boldsymbol{s}[n] = (\boldsymbol{q}[n], \boldsymbol{q}_{E}^{xy}[n])$. Specifically, $\boldsymbol{q}[n]$ represents the positions of all UAVs at time slot $n$, and $\boldsymbol{q}_{E}^{xy}[n]$ is the coordinates of the eavesdroppers within the $x$-$y$ plane at time slot $n$.

\subsubsection{\textbf{Action Space}}
\label{SubSubSection: Action Space}

\par At a certain time slot $n$, each UAV needs to choose its own proper excitation current weight and position. Accordingly, the action set of UAV swarm can be represented by $\boldsymbol{a}[n] = (\boldsymbol{I}[n], \boldsymbol{q}[n])$, where $\boldsymbol{I}[n]$ and $\boldsymbol{q}[n]$ represent the excitation current weights and positions of all UAVs at time slot $n$, respectively.

\subsubsection{\textbf{Reward Function}}
\label{SubSubSection: Reward Function}

\par In DRL, the reward garnered from the agent-environment interchange provides a quantifiable measure of action efficiency in a given state. Therefore, the formulated ASCEE-MOP can be transformed into maximizing the accumulative reward. Accordingly, the reward function can be constructed as follows:
\begin{equation}
	r[n] = \omega_{1}r_{SE}[n] + \omega_{2}r_{E}[n] - r_{P}[n],
	\label{Equation: reward design}
\end{equation}

\noindent where the first term, i.e., $r_{SE}[n] = R_{SE}[n]$ represents the secrecy rate that the system achieves at time slot $n$. Moreover, the second term $r_{E}[n] = -E[n]$ quantifies the total flight energy consumption of all UAVs at time slot $n$. Furthermore, $\omega_{1}$ and $\omega_{2}$ denote the weight factors for the two objectives, which can be determined based on their respective value ranges. In addition, the penalty $r_{P}[n]$ is applied if the UAVs violate the constraint of speed or collide with each other.

\subsubsection{\textbf{Transition Probability}}
\label{SubSubSection: Transition Probability}

\par In our work, the transition probability of the state, which is denoted as $\mathcal{P}(\boldsymbol{s}[n+1]|\boldsymbol{s}[n], \boldsymbol{a}[n])$, specifies the probability distribution of the subsequent state after the UAVs execute their respective actions in the current state.

\subsection{Basic Principles of Conventional TD3}
\label{SubSection: Basic Principles of Conventional TD3}

\par TD3\cite{Fujimoto2018} is an advanced reinforcement learning algorithm that extends from the foundations of deep deterministic policy gradient (DDPG)\cite{Lillicrap2016} method. Specifically, TD3 addresses the key limitations in DDPG by incorporating several novel techniques including twin critic networks, delayed policy updates, and target policy smoothing, which collectively contribute to its superior performance in continuous control tasks.

\subsubsection{Actor-Critic Framework}
\label{SubSubSection: Actor-Critic Framework}
Similar to DDPG, TD3 employs an actor-critic structure, where the actor network $\boldsymbol{\mu}(\boldsymbol{s}|\boldsymbol{\theta_{\mu}})$ outputs deterministic actions, and the critic networks $\boldsymbol{Q}(\boldsymbol{s},\boldsymbol{a}|\boldsymbol{\theta_{Q}})$ evaluate the action-state value function. The objective is to find the optimal policy $\boldsymbol{\pi}$ that maximizes the expected accumulated return.

\par The Bellman equation provides a recursive decomposition to update the action-value function $\boldsymbol{Q}(\boldsymbol{s}, \boldsymbol{a})$, which can be described mathematically as follows \cite{Sutton2018reinforcement}:
\begin{equation}
    \begin{aligned}
        \boldsymbol{Q}(\boldsymbol{s}[n], \boldsymbol{a}[n]) = r[n] + \gamma \mathbb{E}_{\boldsymbol{s}[n+1] \sim \boldsymbol{p_{\pi}}}[&\boldsymbol{Q}(\boldsymbol{s}[n+1], \\&\boldsymbol{\mu}(\boldsymbol{s}[n+1]))],
    \end{aligned}
    \label{Equation: Bellman equation}
\end{equation}

\noindent where $\boldsymbol{p_{\pi}}$ represents the transition probability distribution under policy $\boldsymbol{\pi}$.

\subsubsection{Twin Critic Networks}
\label{SubSubSection: Twin Critic Networks}

\par One of the significant improvements in TD3 is the use of twin critic networks to address overestimation bias. Specifically, overestimation usually occurs when the action-value estimates are consistently higher than the true values, leading to the suboptimal policy updates. While in TD3, two independent critic networks, i.e., $\boldsymbol{Q_1}(\boldsymbol{s}, \boldsymbol{a}|\boldsymbol{\theta_{Q_1}})$ and $\boldsymbol{Q_2}(\boldsymbol{s}, \boldsymbol{a}|\boldsymbol{\theta_{Q_2}})$, are used to estimate the value of state-action pairs. The target Q-value is computed as the minimum of the two estimates, which is represented as follows:
\begin{equation}
	\label{Equation: Target Q-value}
	y[n] = r[n] + \gamma \min_{i=1,2} \boldsymbol{Q^\prime_i}(\boldsymbol{s}[n+1], \boldsymbol{\mu^\prime}(\boldsymbol{s}[n+1]|\boldsymbol{\theta_{\mu}^{\prime}})),
\end{equation}

\noindent where $\boldsymbol{Q^\prime_i}$ is the target critic networks corresponding to $\boldsymbol{Q_i}$, and $\boldsymbol{\mu^\prime}$ is the target actor network.

\subsubsection{Delayed Policy Update}
\label{SubSubSection: Delayed Policy Update}

\par TD3 incorporates the delayed policy update to prevent the policy network from overfitting to noisy value estimates. While the critic networks are updated at each time step, the actor network is updated less frequently. Specifically, the policy is updated every $d$ iterations of the critic networks, and this delay allows the value estimates to stabilize, leading to more reliable policy updates.

\subsubsection{Target Policy Smoothing}
\label{SubSubSection: Target Policy Smoothing}

\par To further enhance the stability, TD3 introduces target policy smoothing, which adds extra noise to the target action during the critic update process. This process involves sampling noise from a Gaussian distribution $\boldsymbol{\epsilon} \sim \mathcal{N}(0, \sigma^{2})$ and clipping it to a certain range to maintain the target action within the permissible action space. Specifically, the process above can be represented as follows:
\begin{equation}
	\label{Equation: Target Policy Smoothing}
	\boldsymbol{sa}[n+1] = \boldsymbol{\mu^\prime}(\boldsymbol{s}[n+1]|\boldsymbol{\theta_{\mu}^{\prime}}) + \boldsymbol{\epsilon}, \boldsymbol{\epsilon} \sim \text{clip}(\mathcal{N}(0, \sigma^{2}), -c, c),
\end{equation}

\noindent where $\text{clip}(x, a, b)$ is a clipping operator, which is defined as $\text{clip}(x, a, b) = x$ if $a < x < b$, $\text{clip}(x, a, b) = a$ if $x \leq a$ and $\text{clip}(x, a, b) = b$ if $x \geq b$. This smoothed target action $\boldsymbol{sa}[n+1]$ is used in the Bellman update to replace the target action $\boldsymbol{\mu^\prime}(\boldsymbol{s}[n+1]|\boldsymbol{\theta_{\mu}^{\prime}})$ in Eq.~\eqref{Equation: Target Q-value}, which reduces the variance of the value estimates and preventing sharp changes in the policy.

\subsubsection{Network Training}
\label{SubSubSection: Network Training}

\par The training process of TD3 involves updating the actor and critic networks based on specific loss functions, which is designed to improve the learning stability and performance. The update of critic network is through minimizing the temporal difference (TD) error loss function, which is defined as follows:
\begin{equation}
	\label{Equation: Critic Loss}
	L(\boldsymbol{\theta_{Q_i}}) = \mathbb{E} \left[ \left( \boldsymbol{Q_i}(\boldsymbol{s}[n], \boldsymbol{a}[n]|\boldsymbol{\theta_{Q_i}}) - y[n] \right)^2 \right], i=1,2.
\end{equation}

\par With a batch of randomly sampled $B$ transitions from experience replay buffer $\mathcal{D}$, the loss function for the critic network can be approximated as follows:
\begin{equation}
	\label{Equation: Estimated Critic Loss}
	L(\boldsymbol{\theta_{Q_i}}) \approx \frac{1}{B} \sum_{b=1}^B \left(\boldsymbol{Q_i}(\boldsymbol{s_b}, \boldsymbol{a_b}|\boldsymbol{\theta_{Q_i}}) - y_b \right)^2, i=1,2,
\end{equation}

\noindent where $y_b = r_b + \gamma \min_{i=1,2} \boldsymbol{Q^\prime_i}(\boldsymbol{s\__{b}}, \boldsymbol{\mu^\prime}(\boldsymbol{s\__{b}}|\boldsymbol{\theta_{\mu}^{\prime}}) + \boldsymbol{\epsilon})$.

\par The actor network $\boldsymbol{\mu}(\boldsymbol{s}|\boldsymbol{\theta_\mu})$ is updated less frequently than the critic networks to ensure stable learning. The objective of actor network is to maximize the expected Q-value as evaluated by the first critic network. The loss function for the actor network is represented as follows:
\begin{equation}
	\label{Equation: Actor Loss}
	L(\boldsymbol{\theta_\mu}) = -\mathbb{E} \left[\boldsymbol{Q_1}(\boldsymbol{s}, \boldsymbol{\mu}(\boldsymbol{s}|\boldsymbol{\theta_\mu})|\boldsymbol{\theta_{Q_1}}) \right].
\end{equation}

\par With a batch of randomly sampled $B$ transitions from experience replay buffer $\mathcal{D}$, the loss function for the actor network can be approximated as follows:  
\begin{equation}
	\label{Equation: Estimated Actor Loss}
	L(\boldsymbol{\theta_\mu}) \approx -\frac{1}{B} \sum_{b=1}^B \boldsymbol{Q_1}(\boldsymbol{s_b}, \boldsymbol{\mu}(\boldsymbol{s_b}|\boldsymbol{\theta_\mu})|\boldsymbol{\theta_{Q_1}}).
\end{equation}

\par The target networks are updated using a soft update mechanism, which blends the parameters of the main networks with those of the target networks using a weight factor. The updates are defined as follows:
\begin{equation}
	\label{Equation: Soft Critic}
	\boldsymbol{\theta_{Q_i}^{\prime}} \leftarrow \tau \boldsymbol{\theta_{Q_i}} + (1 - \tau) \boldsymbol{\theta_{Q_i}^{\prime}}, i=1,2,
\end{equation}
and 
\begin{equation}
	\label{Equation: Soft Actor}
	\boldsymbol{\theta_{\mu}^{\prime}} \leftarrow \tau \boldsymbol{\theta_{\mu}} + (1 - \tau) \boldsymbol{\theta_{\mu}^{\prime}},
\end{equation}

\noindent where \(\tau\) is a small soft weight factor. It can be observed that the updated parameters of a target network are a weighted combination of its original parameters and the corresponding network parameters.

\subsection{Generative Diffusion Model for Actor Network}
\label{SubSection:Generative Diffusion Model for Actor Network}

\par In this section, we first elaborate the motivation behind employing diffusion models within the actor network of TD3 algorithm. Then, we explore the customization of the diffusion model for generating optimal decisions regarding the formulated ASCEE-MOP.

\subsubsection{Motivation of Employing Diffusion Model}
\label{SubSubSection: Motivation of Adopting Diffusion Model}

\par Deep reinforcement learning (DRL) has become an effective method for dealing with various network optimization problems in dynamic environments. Generally, DRL employs deep neural networks (DNNs) to provide optimal actions according to the current environment state. Multi-layer perceptrons (MLPs), a prevalent fully-connected DNN architecture in DRL, consist of hidden layers with nonlinear activation functions. However, the ASCEE-MOP faces unique challenges, such as the mobility of eavesdroppers, which introduces uncertainty and results in a highly dynamic and complex state space. Moreover, ASCEE-MOP involves intricate trade-offs between various optimization objectives, making it challenging to identify optimal solutions in this constantly changing environment. Thus, traditional MLP approaches may struggle to fully capture and balance these interconnected objectives.

\par In contrast, generative diffusion models \cite{Harshvardhan2020}, \cite{Yang2024}, with their superior feature learning capabilities, can better comprehend environmental states and the relationships between different objectives. This understanding allows DRL agents to make more balanced and optimized decisions in the highly uncertain and dynamic environment of ASCEE-MOP. Consequently, the use of diffusion models can be highly advantageous for addressing the complex issues inherent in ASCEE-MOP.

\subsubsection{Diffusion Model}
\label{SubSubSection: Diffusion Model}

\par Diffusion model, such as the denoising diffusion probabilistic model (DDPM)\cite{Ho2020}, operate through a dual-phase process that are the forward process and reverse process. Specifically, the forward phase incrementally adds Gaussian noise to the data, converting it progressively into a pure noise distribution. Conversely, the reverse phase reconstructs the original data by systematically removing this noise.

\par \textbf{Forward Process:} Given a original data $\bm{x}_0$, the forward process produces a series of noisy samples $\{\bm{x}_t\}_{t=0}^T$ by gradually adding the Gaussian noise. Specifically, at each step $t$, the noisy sample $\bm{x}_t$ is sampled from the distribution $\boldsymbol{p}(\bm{x}_t | \bm{x}_{t-1})$, which is generated from the previous sample $\bm{x}_{t-1}$ by using the method as follows:
\begin{equation}
	\label{Equation: Forward Process}
	\boldsymbol{p}(\bm{x}_t | \bm{x}_{t-1}) = \mathcal{N}(\bm{x}_t; \sqrt{1-\beta_t} \bm{x}_{t-1}, \beta_t \boldsymbol{I}),
\end{equation}

\noindent where $\boldsymbol{I}$ represents the identity matrix, and $\beta_t$ is a variance schedule that is controlled by the variance preserving (VP) schedule. Moreover, $\beta_t$ is the variance function of VP
stochastic differential equations, which is as follows \cite{Xiao2022}:
\begin{equation}
	\label{Equation: Beta Schedule}
	\beta_t = 1 - e^{-\frac{\beta_{\text{min}}}{T} - \frac{2t-1}{2T^{2}} (\beta_{\text{max}} - \beta_{\text{min}})},
\end{equation}

\noindent where $\beta_{min}$ and $\beta_{max}$ are the two constants that define the minimum and maximum variance.

\par The entire forward process from $\bm{x}_0$ to $\bm{x}_T$ can be expressed as follows:
\begin{equation}
	\label{Equation: Forward Process Full}
	\boldsymbol{p}(\bm{x}_{T} | \bm{x}_0) = \prod_{t=1}^T \boldsymbol{p}(\bm{x}_t | \bm{x}_{t-1}).
\end{equation}

\par Moreover, the forward process that delineates the mathematical relation between $\bm{x}_0$ and any $\bm{x}_t$ is described as follows:
\begin{equation}
	\label{Equation: Relation between x0 and xt}
	\bm{x}_t =\sqrt{\bar{\alpha}_t}\boldsymbol{x}_0+\sqrt{1-\bar{\alpha}_t}\boldsymbol{\epsilon},
\end{equation}

\noindent where $\bar{\alpha}_t = \prod_{k=1}^t \alpha_k$ represents the cumulative product of $\alpha_k$ for all steps $k \leq t$, wherein $\alpha_t = 1 - \beta_t$, and $\boldsymbol{\epsilon}\sim\mathcal{N}\left(\mathbf{0},\mathbf{I}\right)$ is a standard Gaussian noise. With an increase in $t$, $\bm{x}_T$ gradually transitions into purely noise, adhering to an isotropic Gaussian distribution $\mathcal{N}(0, \boldsymbol{I})$. However, note that due to the absence of an optimal decision solution dataset (i.e., $\bm{x}_0$ in the forward process) for the formulated optimization problem, the forward process is not integrated into the proposed GDMTD3.

\par \textbf{Reverse Process:} In the reverse process, the goal is to recover the original data $\bm{x}_0$ from a noisy sample $\bm{x}_T$ that follows a standard Gaussian distribution $\mathcal{N}(\bm{0}, \boldsymbol{I})$ by iteratively removing the noise. However, the statistical distribution $q(\bm{x}_{t-1} | \bm{x}_t)$ necessitate computations that involve the data distribution, which is typically intractable in practice. Instead, our strategy is to approximate the conditional distribution $q(\bm{x}_{t-1} | \bm{x}_t)$ by using a parameterized model $p_{\theta_{d}}$, which can be expressed as follows:
\begin{equation}
	\label{Equation: Approximate Reverse Process}
	\boldsymbol{p_{\theta_{d}}}(\bm{x}_{t-1} | \bm{x}_t) = \mathcal{N}(\bm{x}_{t-1}; \boldsymbol{\kappa_{\theta_{d}}}(\bm{x}_t, t, \bm{g}), \tilde{\beta}_t \boldsymbol{I}),
\end{equation}

\noindent where $\boldsymbol{\kappa_{\theta_{d}}}(\bm{x}_t, t, \bm{g})$ is the mean, wherein $\bm{g}$ is the condition information, and $\tilde{\beta}_t$ represents a predetermined variance factor, which is represented as follows:
\begin{equation}
	\label{Equation: Predetermined Variance}
	\tilde{\beta}_t = \frac{1 - \bar{\alpha}_{t-1}}{1 - \bar{\alpha}_t} \beta_t.
\end{equation}

\par Utilizing Bayesian formulation, the reverse process is restructured as a Gaussian probability density function. The mean for the reverse process is computed as follows \cite{Ho2020}:
\begin{equation}
	\label{Equation: Reverse Mean}
	\boldsymbol{\kappa_{\theta_{d}}}(\bm{x}_t, t, \boldsymbol{g}) = \frac{\sqrt{\alpha_t}\left(1-\bar{\alpha}_{t-1}\right)}{1-\bar{\alpha}_t}\bm{x}_t+\frac{\sqrt{\bar{\alpha}_{t-1}}\beta_t}{1-\bar{\alpha}_t}\bm{x}_0.
\end{equation}

\par Nonetheless, the parameterized model $\boldsymbol{p_{\theta_{d}}}$ does not have access to $\bm{x}_0$ and therefore must estimate it as a substitute. According to Eq.~\eqref{Equation: Relation between x0 and xt}, $\bm{x}_0$ can be calculated as follows:
\begin{equation}
	\label{Equation: Derived x_0}
	\bm{x}_0 = \frac{1}{\sqrt{\bar{\alpha}_t}} \left( \bm{x}_t - \sqrt{1 - \bar{\alpha}_t} \cdot \boldsymbol{\varepsilon_{\theta_{d}}}(\bm{x}_t, t, \bm{g})\right),
\end{equation}

\noindent where $\boldsymbol{\varepsilon_{\theta_{d}}}(\bm{x}_t, t, \bm{g})$ is a deep neural network that generates the denoising noise based on the condition $\bm{g}$, and then indirectly approximate the mean by
\begin{equation}
	\label{Equation: Approximate Mean}
	\boldsymbol{\kappa_{\theta_{d}}}(\bm{x}_t, t, \bm{g}) = \frac{1}{\sqrt{\alpha_t}} \left( \bm{x}_t - \frac{\beta_t \cdot \boldsymbol{\varepsilon_{\theta_{d}}}(\bm{x}_t, t, \bm{g})}{\sqrt{1 - \bar{\alpha}_t}} \right).
\end{equation}

\par Tracing the reverse transitions from $\bm{x}_T$ back to $\bm{x}_1$, we can establish the generative distribution $\boldsymbol{p_{\theta_{d}}}(\bm{x}_0)$ as follows:
\begin{equation}
	\label{Equation: Generative Distribution}
	\boldsymbol{p_{\theta_{d}}}(\bm{x}_0) = \boldsymbol{p}(\bm{x}_T) \prod_{t=1}^T \boldsymbol{p_{\theta_{d}}}(\bm{x}_{t-1} | \bm{x}_t),
\end{equation}

\noindent where $p(\bm{x}_T)$ represents a standard normal distribution. Once the generative distribution $\boldsymbol{p_{\theta_{d}}}(\bm{x}_0)$ is successfully trained, we can then proceed to sample $\bm{x}_0$ from Eq.~\eqref{Equation: Generative Distribution}.

\begin{figure}[!t]
	\removelatexerror
	\begin{algorithm}[H]
		\raggedright
		\caption{Action Sampling Based on Generative Diffusion Model}
		\label{Algorithm 1: Generative Diffusion Model Sample Actions}
		\LinesNumbered
		\KwIn{The state of current environment $\boldsymbol{s}[n]$}
            \KwOut{The action decision $\boldsymbol{a}[n]$}
				Initialize a random Gaussian distribution $\bm{x}_{T} \sim \mathcal{N}(0, \bm{I})$; \\
				\For{\rm{the denoising step} $t = T$ \KwTo $1$}{
					Deduce a denoising distribution $\boldsymbol{\varepsilon_{\theta_{d}}}(\bm{x}_t, t, \boldsymbol{s}[n])$ by a deep neural network;\\
					Compute the mean $\boldsymbol{\kappa_{\theta_{d}}}(\bm{x_{t}}, t, \boldsymbol{s}[n])$ of $\boldsymbol{p_{\theta_{d}}}(\bm{x}_{t-1}|\bm{x}_{t})$ according to Eq.~\eqref{Equation: Approximate Mean};\\
					Compute the distribution $\bm{x}_{t-1}$ using the reparameterization trick according to Eq.~\eqref{Equation: reparameterization technique};
				}
				Compute the distribution of $\bm{x}_{0}$ according to Eq.~\eqref{Equation: Generative Distribution} and randomly select an action $\boldsymbol{a}[n]$ based on it; \\
        \Return $\boldsymbol{a}[n]$
	\end{algorithm}
\end{figure}

\subsubsection{Integration of Diffusion Model and Actor Network of TD3}
\label{SubSubSection: Integration with Actor Network}

\par Integrating diffusion model into the actor network of conventional TD3 algorithm significantly enhances the decision-making by providing a more diverse set of potential actions. Specifically, the generative capabilities of diffusion model allow for the creation of complex action sets, which are refined through the learned reverse process, enabling direct sampling of actions from the generative distribution $\boldsymbol{p_{\theta_{d}}}(\bm{x}_0)$.

\par A significant challenge in integrating  diffusion model is managing stochastic components, which complicates gradient descent methods typically used in training. To overcome this issue, a reparameterization process that facilitates differentiable sampling is employed, which can be represented as follows:
\begin{equation}
	\label{Equation: reparameterization technique}
	\boldsymbol{x}_{t-1}=\boldsymbol{\kappa}_{\boldsymbol{{\theta_{d}}}}\left(\boldsymbol{x}_t,t,\bm{s}\right)+\left(\tilde{\beta}_t/2\right)^2\odot\boldsymbol{\epsilon},
\end{equation}

\noindent where $\bm{s}$ which represents the current state of the environment in DRL, is used as a conditional variable in the parameterization function $\boldsymbol{\kappa}_{\boldsymbol{{\theta_{d}}}}$. Moreover, $\odot$ is the operator of Hadamard product. 

\par This adaptation allows the diffusion process to be contextually responsive and adjusting actions dynamically according to the state of the environment, which is crucial for DRL algorithms where the environmental state guides the necessary action responses. Accordingly, the main steps of the action sampling process based on generative diffusion model is detailed in Algorithm \ref{Algorithm 1: Generative Diffusion Model Sample Actions}. 

\subsection{Main Flow of Proposed Algorithm}
\label{SubSection: Flow and Analysis of the Proposed Algorithm}

\par
\begin{figure*}[htbp]
	\centering	\includegraphics[width=\linewidth,scale=1.00]{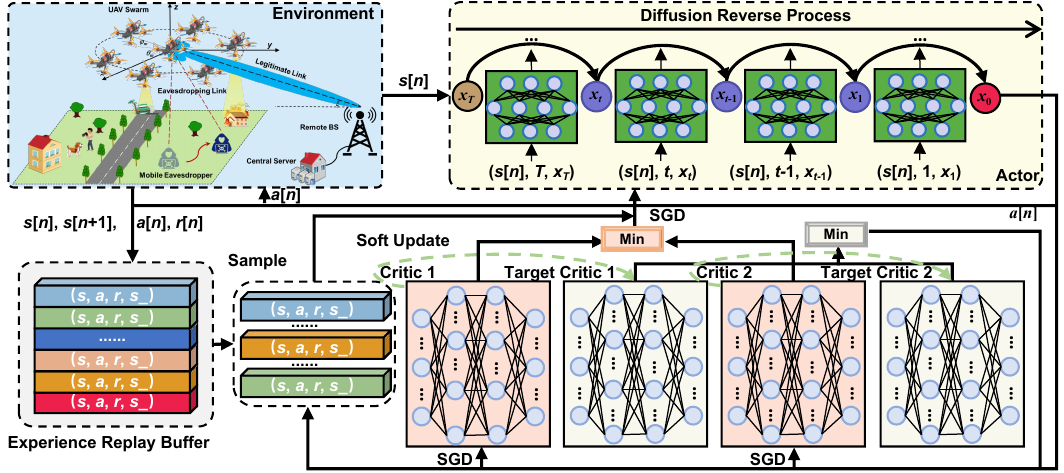}
	\caption{Schematic of GDMTD3 framework, where the generative diffusion model is integrated into the actor network of TD3 algorithm to capture complex state features and generate optimal actions according to the current state of the environment.}
	\label{Figure:Algorithm Framework}
\end{figure*}

 Fig.~\ref{Figure:Algorithm Framework} shows the framework and main flow of the proposed GDMTD3 for the formulated ASCEE-MOP. Specifically, the proposed method integrates the diffusion model within DRL, which enhances the capability of the actor network for navigating the complex decision spaces under high-dimensional and noisy input data. The detailed implementation of this process is elaborated in Algorithm~\ref{Algorithm 2:GDMTD3}.

\begin{figure}[!t]
	\removelatexerror
	\begin{algorithm}[H]
		\raggedright
		\caption{GDMTD3}
		\label{Algorithm 2:GDMTD3}
		\LinesNumbered
		Initialize two online critic networks denoted as $\boldsymbol{Q_{1}}$ and $\boldsymbol{Q_{2}}$ with parameters $\boldsymbol{\theta_{Q_{1}}}$ and $\boldsymbol{\theta_{Q_{2}}}$ and a generative diffusion-enabled online actor network denoted as $\boldsymbol{\varepsilon}$ with parameters $\boldsymbol{\theta_{d}}$; \\
		Initialize the corresponding target networks: $\boldsymbol{\theta_{Q_1}^{\prime}} \gets \boldsymbol{\theta_{Q_{1}}}, \boldsymbol{\theta_{Q_2}^{\prime}} \gets \boldsymbol{\theta_{Q_{2}}}$ and $\boldsymbol{\theta_{\mu}^{\prime}} \gets \boldsymbol{\theta_{\mu}}$; \\
		\For{\rm{the training episode} = $1$ \KwTo $M$}{
			Reset the initial state $\boldsymbol{s}[0]$ of environment;\\
			\Repeat{environment is terminated}{
				$step \gets 0$;\\
				Call \textbf{Algorithm~\ref{Algorithm 1: Generative Diffusion Model Sample Actions}} to obtain the action $\boldsymbol{a}[step]$;\\
				Execute the action $\boldsymbol{a}[step]$ in the environment and receive the reward $r[step]$ and the next state $\boldsymbol{s}[step+1]$ from the environment;\\
				Store the experience $(\boldsymbol{s}[step], \boldsymbol{a}[step], r[step], \boldsymbol{s}[step+1])$ in the replay buffer $\mathcal{D}$;\\
				Sample a random batch $\mathcal{B}$ from the replay buffer $\mathcal{D}$;\\
				Update the online critic network parameters according to Eq.~\eqref{Equation: Estimated Critic Loss};\\
				\If{$step\mod d$}
				{
					Update the actor network parameters according to Eq.~\eqref{Equation: Estimated Actor Loss};\\
					Soft-update the target networks according to Eqs.~\eqref{Equation: Soft Critic} and \eqref{Equation: Soft Actor};}
				$step \gets step + 1$;
			}
			
		}
	\end{algorithm}
\end{figure}

\subsubsection{\textbf{Training and Execution}}
\label{SubSubSection: Training and Execution}
\par In the considered UAV swarm-enabled surveillance network system, the RBS coordinates the training phase through an actor-critic network framework. In this phase, the interaction information between UAV swarm and the environment is regularly recorded and stored into a replay buffer. Note that the RBS possesses the sufficient capabilities to transmit the training parameters to UAV swarm \cite{Chen2021}. Following a comprehensive training period, the actor network is then integrated with UAV swarm, steering their real-time operations to adaptively accomplish the secure communication mission throughout the execution phase.

\subsubsection{\textbf{Complexity Analysis}}
\label{SubSubSection: Complexity Analysis}
\par In this section, we analyze the computational and space complexity of GDMTD3 during training and execution phases.
\par \textbf{Training Phase:} The computational complexity of GDMTD3 is $\mathcal{O}(4 |\boldsymbol{\theta_{Q_1}}| + 2 |\boldsymbol{\theta_{d}}| + MNT|\boldsymbol{\theta_{d}}| + MNV + MN (2 |\boldsymbol{\theta_{Q_1}}|) + MN/d (2 |\boldsymbol{\theta_{Q_1}}| + 2 |\boldsymbol{\theta_{d}}|))$ in the training phase, which can be summarized as follows:
\begin{itemize}
	\item \textbf{\textit{Network Initialize}}: This phase involves the initialization of network parameters. Specifically, the computational complexity is expressed as $\mathcal{O}(4 |\boldsymbol{\theta_{Q_1}}| + 2 |\boldsymbol{\theta_{d}}|)$, where $|\boldsymbol{\theta_{Q_1}}|$ denotes the number of parameters in each of the twin online critic networks, and $|\boldsymbol{\theta_{d}}|$ represents the number of parameters in the diffusion-enabled online actor network.
	
	\item \textbf{\textit{Action Sampling}}: This phase entails generating actions according to the current state using the diffusion reverse process, and its complexity is $\mathcal{O}(MNT|\boldsymbol{\theta_{d}}|)$. Here, $M$ denotes the number of training episodes, $N$ is the number of steps per episode, and $T$ is the number of denoising steps required to sample an action in diffusion-enabled actor network.
	
	\item \textbf{\textit{Replay Buffer Collection:}} The complexity of collecting state transitions in the replay buffer is $\mathcal{O}(MNV)$, where $V$ represents the complexity of interacting with environment.
	
	\item \textbf{\textit{Network Update:}} The updating phase is divided into three main parts that are the frequent updates of the critic networks and less frequent updates of the actor network along with their respective soft updates. Thus, the complexity for this phase is calculated as $\mathcal{O}(MN (2 |\boldsymbol{\theta_{Q_1}}|) + MN/d (2 |\boldsymbol{\theta_{Q_1}}| + 2 |\boldsymbol{\theta_{d}}|))$.
\end{itemize}

\par In the training phase, the space complexity of GDMTD3 is $\mathcal{O}(4|\boldsymbol{\theta_{Q_1}}|+2|\boldsymbol{\theta_{d}}|)+D\left(2|\boldsymbol{s}|+|\boldsymbol{a}|+1\right))$, where $D$ represents the size of the replay buffer and $|\boldsymbol{s}|$, $|\boldsymbol{a}|$ denote the dimensions of the state and action spaces, respectively. This space complexity accounts for the storage of neural network parameters and the data structures required to maintain the replay buffer, which holds tuples of states, actions, rewards, and next states.

\par \textbf{Execution Phase:} During the execution phase, the computational complexity of GDMTD3 is $\mathcal{O}(MNT|\boldsymbol{\theta_{d}}|)$, which can be contributed by action selection according to the current state using the diffusion-enabled actor network. Moreover, the space complexity during the execution phase is $\mathcal{O}(|\boldsymbol{\theta_{d}}|)$ since the diffusion-enabled actor network parameters need to be stored in memory for action selection.

\section{Simulation Results}
\label{Section:Simulation Results}

\par In this section, we present the comprehensive evaluations of our proposed approach and verify the effectiveness and robustness of the proposed GDMTD3 in addressing ASCEE-MOP under various settings.

\subsection{Simulation Setup}
\label{SubSection:Simulation Setup}

\par This section provides an extensive description of the simulation setup, including the simulation platform, environmental details, model design, and benchmarks utilized to evaluate the performance of the proposed approach.

\subsubsection{Simulation Platform}
\label{SubSubSection: Simulation Platform}

\par Our experiments are conducted using a computing setup that included an NVIDIA GeForce RTX 3090 GPU with 24 GB of memory and a 13th Gen Intel(R) Core(TM) i9-13900K 32-core processor with 128 GB of RAM. The operating system on the workstation is Ubuntu 22.04.3 LTS. For our deep learning computations, we use PyTorch 2.2.2, along with the CUDA 11.8.

\subsubsection{Environmental Details}
\label{SubSubSection: Environmental Details}

\par In this study, we consider a UAV swarm consisting of 8 individual UAVs, each of which equipped with a transmit power of $0.1$ W. Moreover, the swarm is dispersed randomly within an area measuring $40$ m by $40$ m. To simulate potential security threats, we incorporate a mobile eavesdropper, which follows the Gauss-Markov mobility model \cite{He2017}. This model is characterized by an average speed of $5.0$ m/s, a correlation coefficient of $0.1$, and a random variance of $1.0$, which together dictate the stochastic and dynamic aspects of the eavesdropper movement. In addition, Table \ref{Table:Other Simulation Parameter Settings} provides the details about the channel characteristics and the UAVs.

\begin{table}
	\renewcommand\arraystretch{1.2}
	\centering 
	\caption{Other Environmental Parameter Settings \cite{Zeng2019} \cite{Yaliniz2016}}
	\label{Table:Other Simulation Parameter Settings}
	
	\begin{tabular}{cc|cc}\toprule[1.5pt]
		\textbf{Parameter} & \textbf{Value} & \textbf{Parameter} & \textbf{Value} \\ \toprule[1pt]
		$f_{c}$ & $2.4$ GHz & $\mu_1$ & $1$ dB \\
		$c_{0}$ & $9.61$ & $\mu_2$ & $20$ dB \\
		$c_{1}$ & $0.16$ & $W$ & $19.6$ N \\
		$v_{0}$ & $4.03$ & $u_{\text{tips}}$ & $120$ \\
		$d_{0}$ & $0.6$ & $\rho$ & $1.225$ \\
		$s$ & $0.05$ & $A$ & $0.503$ \\
		$M$ & $0.1$ & $\kappa$ & $0.012$ \\
		$\Omega$ & $300$ & $\Lambda$ & $0.4$ \\
	\bottomrule[1.5pt]
	\end{tabular}
\end{table}

\subsubsection{Model Design}
\label{SubSubSection: Model Design}

\par
\begin{figure}[t]
	\centering	\includegraphics[width=0.95\linewidth,scale=1.00]{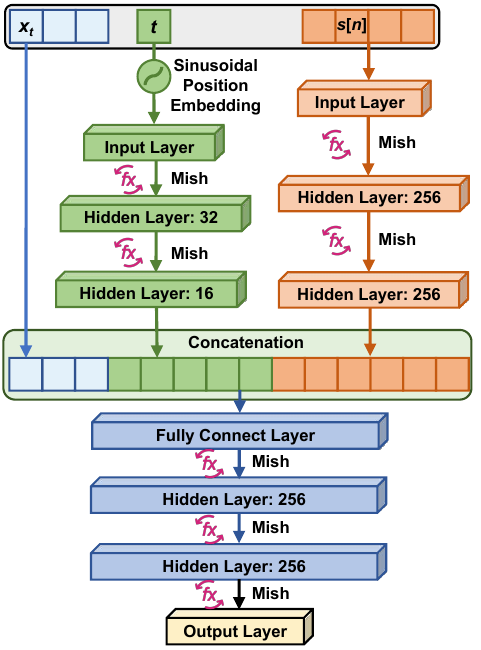}
	\caption{The diffusion-enabled actor network architecture, where Mish activation function \cite{Misra2020} is adopted.}
	\label{Figure:The diffusion-enabled actor network architecture}
\end{figure}

 GDMTD3 utilizes a diffusion model at the core of its actor network, and it employs two structurally identical critic networks to address overestimation issues. Specifically, the critic networks consist of three-layer MLPs with ReLU activation function \cite{Glorot2011}. Moreover, Fig.~\ref{Figure:The diffusion-enabled actor network architecture} shows the detailed configuration of actor network. Specifically, the actor network in GDMTD3 uses sinusoidal position embeddings to capture the temporal dynamics inside the diffusion process and predicts the denoised distribution according to the current state and a random Gaussian distribution. This enhancement enables the actor network to better understand the interdependencies among steps in the diffusion chain. In addition, the Adam optimizer \cite{Kingma2015} is used to train the actor and critic networks, with a learning rate of $lr=3 \times 10^{-4}$ for each network. The target networks, which replicate the structure of the online networks, can minimize the learning variance. We adopt a soft update rate of $\tau=0.005$ as specified in Eqs.~\eqref{Equation: Soft Critic} and \eqref{Equation: Soft Actor}. Additional training hyperparameters are outlined in Table \ref{Table:Other Training Parameter Settings}.

\begin{table}
	\renewcommand\arraystretch{1.2}
	\centering 
	\caption{Other Training Parameter Settings}
	\label{Table:Other Training Parameter Settings}
	
	\begin{tabular}{ccc}\toprule[1.5pt]
		\textbf{Parameter} & \textbf{Description} & \textbf{Value} \\ \toprule[1pt]
		$B$ & Batch size &  $128$ \\
		$\gamma$ & Discount factor & $0.90$ \\
		$D$ & Capacity of the experience replay buffer & $2 \times 10^{6}$\\
		$d$ & Frequency of policy updates & $2$\\
		$T$ & Denoising steps for the diffusion model & $4$ \\
		$M$ & Number of training episodes & $8000$ \\
		\bottomrule[1.5pt]
	\end{tabular}
\end{table}

\subsubsection{Benchmarks}
\label{SubSubSection: Benchmarks}

\par To validate the superiority of our proposed approach, we compare the following approaches:

\begin{itemize}
    \item \textit{\textbf{Random Strategy:}} The random strategy arranges each UAV in a random position within the surveillance area at each time slot, without any specific formation. The excitation current weight for each UAV is also assigned random values within the allowable range. This approach serves as a baseline to evaluate the performance improvements achieved by more strategies.
    \item \textit{\textbf{Linear Antenna Array Strategy:}} The linear antenna array (LAA) strategy arranges UAVs in a linear alignment with an equal inter-UAV separation distance of 0.5 m. Moreover, the geometric center of the linear formation of UAVs coincides with the center of the designated monitoring region.
    \item \textit{\textbf{Planar Antenna Array Strategy:}} The planar antenna array (PAA) strategy arranges UAVs in a two-dimensional grid with an equal inter-UAV separation distance of 0.5 m. Similarly, the geometric center of grid formation of UAVs coincides with the center of the monitoring region.
    \item \textit{\textbf{Circular Antenna Array Strategy:}} The circular antenna array (CAA) strategy arranges UAVs in a circular pattern with a radius of 0.5 m and equal inter-UAV separation distance. Similarly to the LAA and PAA strategies, the center point of this circular UAV formation coincides with the center of the designated monitoring region.
    \item \textit{\textbf{The Proposed GDM-enabled DRL Approach:}} Our approach optimizes the secure rate of system and the flight energy consumption of the UAV swarm by formulating the ASCEE-MOP, and then solving it by using the proposed GDMTD3 algorithm.
\end{itemize}

\par In addition to comparing these approaches, we also compare the proposed GDMTD3 with four well-known DRL benchmarks: DDPG, TD3, SAC \cite{Haarnoja2018}, and PPO \cite{Schulman2017}. Specifically, DDPG, TD3, and SAC are off-policy methods that are used for the continuous action spaces and utilize advanced strategies for stability and performance enhancement. In contrast, PPO is an on-policy method that offers robustness and simplicity in implementation, which is also suitable for the continuous action but focuses on effective policy updates through direct learning from the current policy. Moreover, we implement a transformer-based TD3 method as another point of comparison, which serves as a benchmark to evaluate the capability of the proposed diffusion model in extracting relevant features and representing complex state representations for DRL. Specifically, this method employs a transformer network \cite{Vaswani2017} with two attention heads as the actor network, designed to handle sequential dependencies and complex state representations.

\subsection{Simulation Results}
\label{SubSection:Simulation Results}

\par The detailed results of our simulation are provided in this section. We compare the effectiveness of the proposed GDM-enabled DRL approach with several above-mentioned benchmark deployment policies, and analyze the performance of the proposed GDMTD3 under various algorithm configurations and environmental settings.

\subsubsection{Comparisons with Other Deployment Policies}
\label{SubSubSection: Comparisons with Other Deployment Policies}

\begin{figure}[!t]
	\centering
	\includegraphics[width=\linewidth]{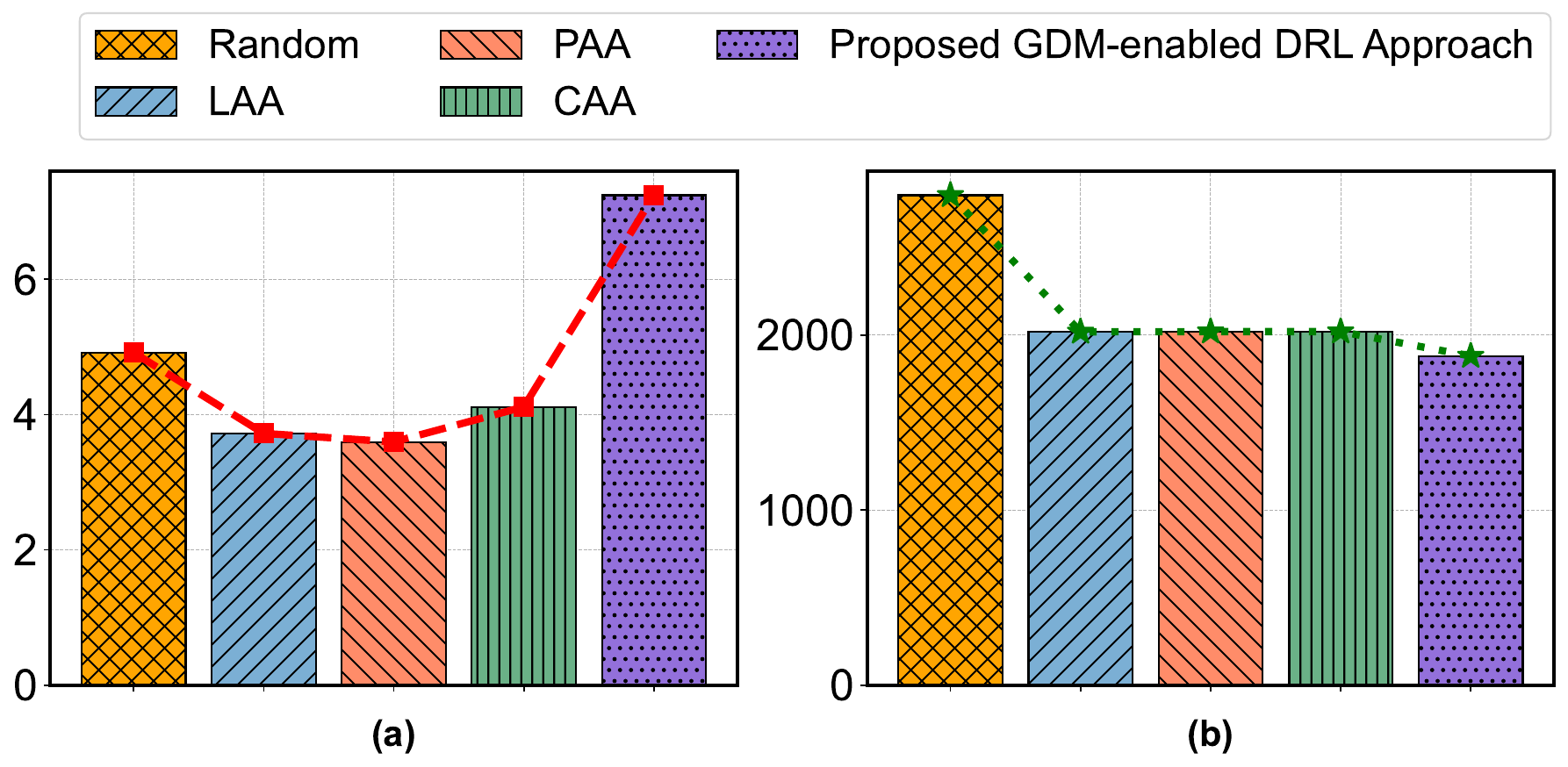}
	\caption{Comparison results of the proposed GDM-enabled DRL approach and other four deployment policies. (a) Average secrecy rate per step. (b) Average flight energy consumption per step.}
	\label{Figure: Optimization results vs Other Deployment Policies}
\end{figure}

\par In this part, the proposed GDM-enabled DRL approach is compared to the four different deployment policies. Specifically, Figs.~\ref{Figure: Optimization results vs Other Deployment Policies}(a) and~\ref{Figure: Optimization results vs Other Deployment Policies}(b) show the average secrecy rate of the system and average flight energy consumption of the UAV swarm, respectively.

\par As shown in Fig.~\ref{Figure: Optimization results vs Other Deployment Policies}(a), the GDM-enabled DRL approach obtains a higher average secrecy rate. This result demonstrates the effectiveness of our proposed approach in ensuring secure communications by optimizing excitation current weights and positions of UAVs. Interestingly, the random strategy performs better than the structured LAA, PAA, and CAA strategies. The most likely reason is that the fixed formations in these three deployment strategies make it more difficult to handle the mobility of the eavesdropper.

\par From Fig.~\ref{Figure: Optimization results vs Other Deployment Policies}(b), it is evident that the suggested GDM-enabled DRL strategy uses less energy on average than the other approaches.
 the proposed GDM-enabled DRL approach exhibits the lower average energy consumption compared to the other strategies. This highlights the efficiency of the proposed GDM-enabled DRL approach in optimizing the flight energy consumption of UAV swarm, which is crucial for the operation of resource-constrained UAVs. Moreover, the random policy shows the highest energy consumption, reflecting its inefficiency. In addition, the LAA, PAA, and CAA strategies demonstrate moderate energy consumption, but they do not achieve the same level of secrecy rate as the proposed GDM-enabled DRL approach, underscoring the advantage of the proposed GDM-enabled DRL approach in optimizing energy consumption while maintaining secure communications.

\par In conclusion, it is apparent that the proposed GDM-enabled DRL approach achieves a superior performance in terms of both the secrecy rate of the system and the flight energy consumption of the UAV swarm.

\subsubsection{Comparisons with Other DRL Benchmarks}
\label{Comparisons with Other DRL Benchmarks}

\begin{figure*}[!t]
    \includegraphics[width=\linewidth]{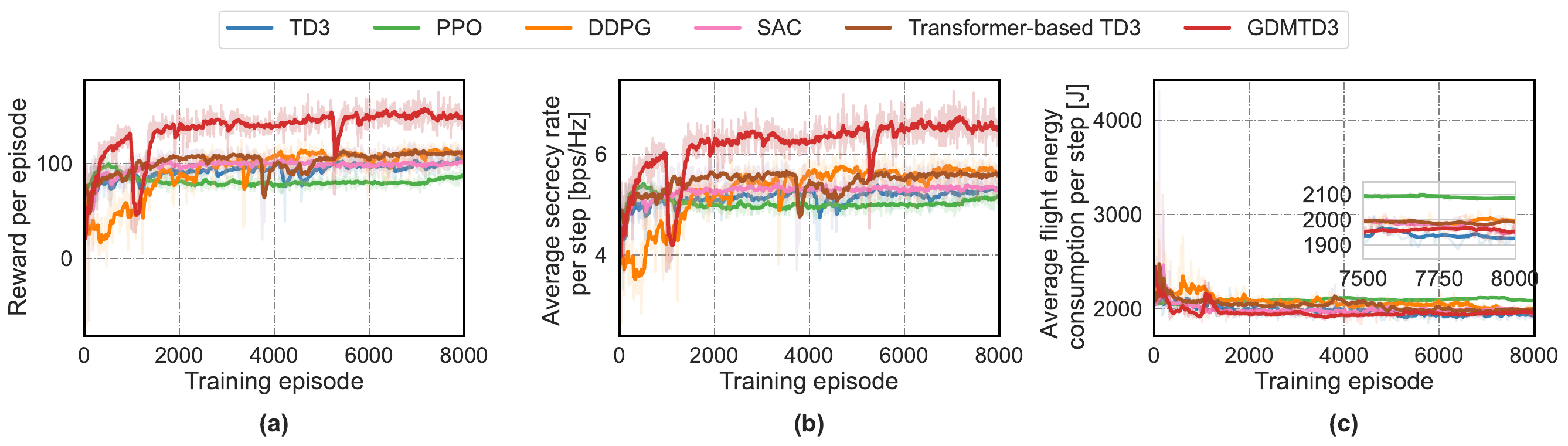}
    \caption{Comparison results of GDMTD3 and DRL benchmarks. (a) Reward per episode. (b) Average secrecy rate per step. (c) Average flight energy consumption per step.}
    \label{Figure: Optimization results vs Other DRL Benchmarks}
\end{figure*}

\par Fig.~\ref{Figure: Optimization results vs Other DRL Benchmarks} shows the comparison results of GDMTD3 with five different DRL benchmarks, including TD3, PPO, DDPG, SAC and transformer-based TD3 methods. As shown in Fig.~\ref{Figure: Optimization results vs Other DRL Benchmarks}(a), the proposed GDMTD3 reports significantly higher rewards per episode than the other DRL methods. This superiority of GDMTD3 is originated from the incorporation of diffusion model in GDMTD3, which allows for more efficient exploration and exploitation of the state-action space, resulting in higher cumulative rewards. Moreover, Figs.~\ref{Figure: Optimization results vs Other DRL Benchmarks}(b) and~\ref{Figure: Optimization results vs Other DRL Benchmarks}(c) indicate that GDMTD3 achieves the highest average secrecy rate of the system and relatively low average flight energy consumption of the UAV swarm among the compared methods. In addition, although the transformer-based TD3 method outperforms traditional TD3, PPO, DDPG, and SAC methods, it does not reach the secrecy rate achieved by GDMTD3, highlighting the advantage of diffusion model in adapting to the complex secure communication scenario involving the mobile eavesdropper.

\subsubsection{Impact of Algorithm Parameters}
\label{SubSubSection: Impact of Algorithm Parameters}

\par In this section, we evaluate effects of different parameters on the performance of GDMTD3 including the random seed, noise schedule function, and denoising step.

\par
\begin{figure}[t]
	\centering	\includegraphics[width=0.9\linewidth,scale=1.00]{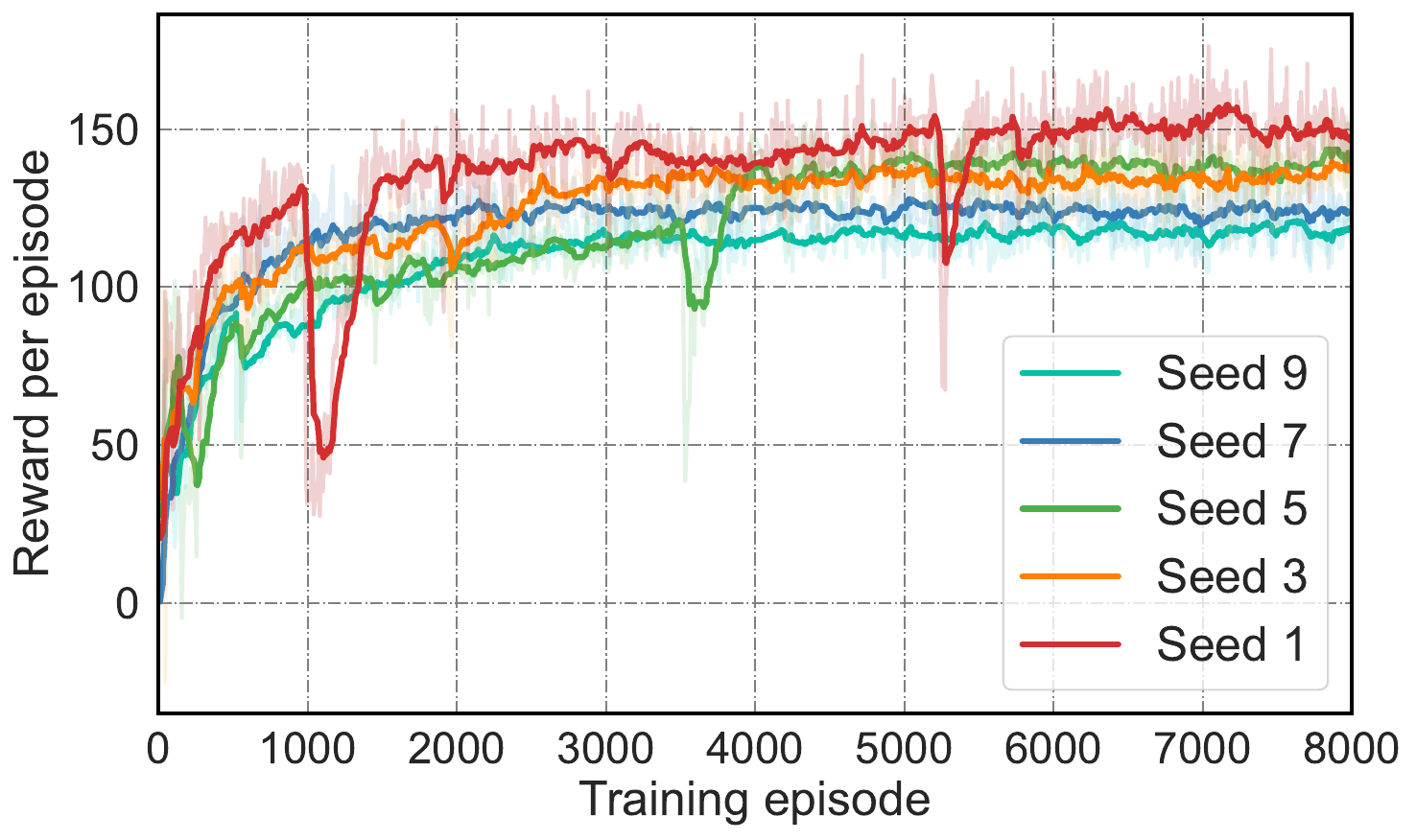}
	\caption{Comparison of reward curves of GDMTD3 with different random seeds.}
	\label{Figure: Comparison of reward curves of GDMTD3 with different random seeds}
\end{figure}

 \textit{\textbf{Effect of Different Random Seeds.}} DRL algorithms are known to be sensitive to random seeds, which can significantly impact their performance, sometimes even causing the algorithm failing to converge when different seeds are used \cite{Colas2018}. Specifically, this sensitivity arises because random seeds influence various aspects of the training process, such as the initialization of neural network weights, the order of data processing, and the exploration strategies. To this end, we compare the impact of different random seeds on the performance of the GDMTD3. As shown in Fig.~\ref{Figure: Comparison of reward curves of GDMTD3 with different random seeds}, GDMTD3 consistently converges and achieves high rewards although the reward curves vary slightly depending on the random seed. This result demonstrates its robustness and stability across different initial conditions.

\par
\begin{figure}[t]
	\centering	\includegraphics[width=0.9\linewidth,scale=1.00]{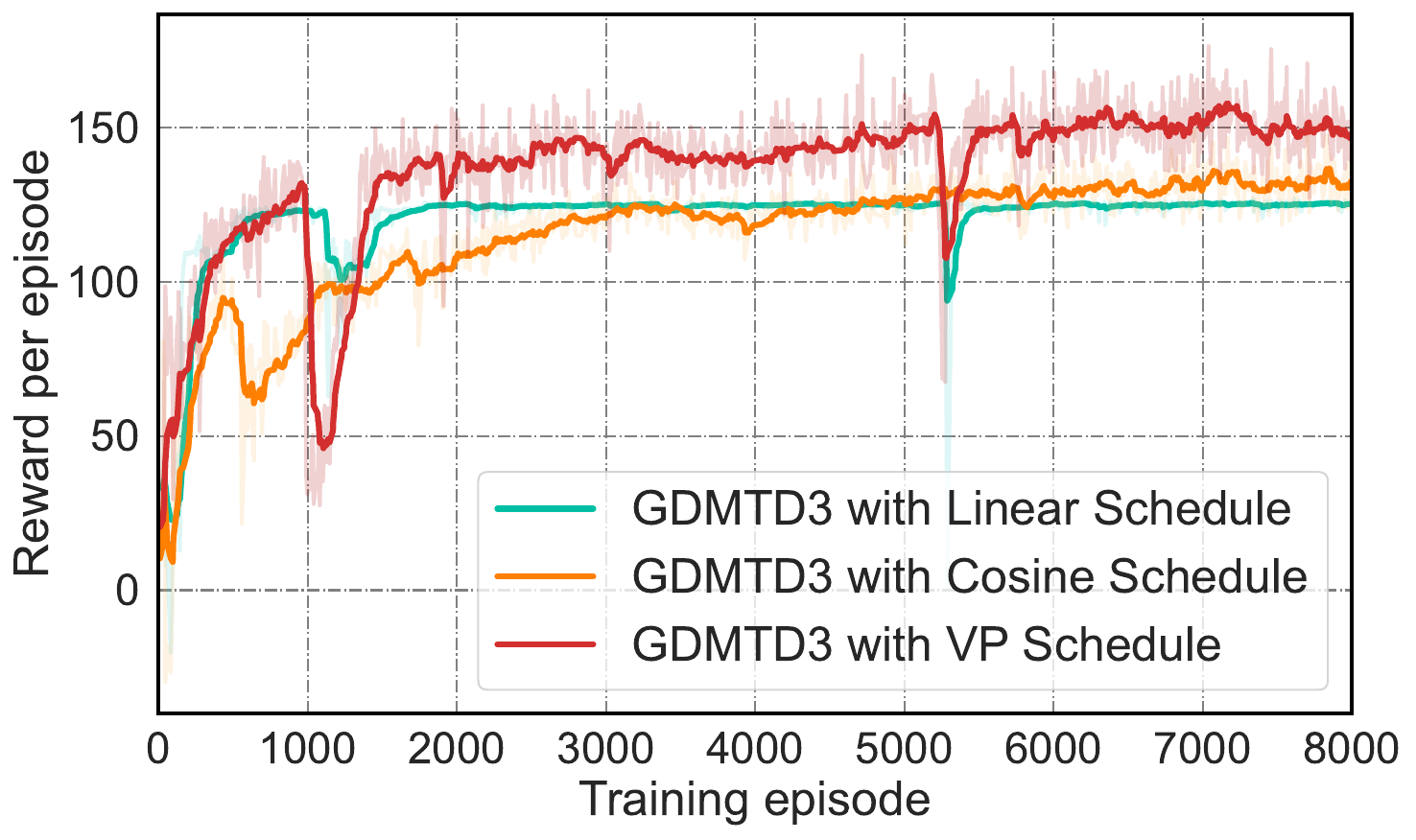}
	\caption{Comparison of reward curves of GDMTD3 with different schedule strategies.}
	\label{Figure: Comparison of reward curves of GDMTD3 with different schedule strategies}
\end{figure}

 \textit{\textbf{Effect of Different Noise Schedule Functions.}} Diffusion-based models are also affected by the selection of noise schedule functions, which determine how parameters such as noise levels are adjusted over time \cite{Nichol2021}. Specifically, this influence stems from the direct effect of noise schedule functions on the diffusion process, which depends on how effectively the model learns to generate high-quality samples. In our scenario, we evaluate the impact of different noise schedule functions on the performance of GDMTD3, which includes VP, linear and cosine noise schedule functions \cite{Nichol2021}. As illustrated in Fig.~\ref{Figure: Comparison of reward curves of GDMTD3 with different schedule strategies}, the results show that the VP schedule leads to the highest reward and faster convergence among the three noise schedule functions. This result highlights the superior performance of the VP schedule when applying GDMTD3 method to address the formulated ASCEE-MOP.

\par
\begin{figure}[t]
	\centering	\includegraphics[width=\linewidth,scale=1.00]{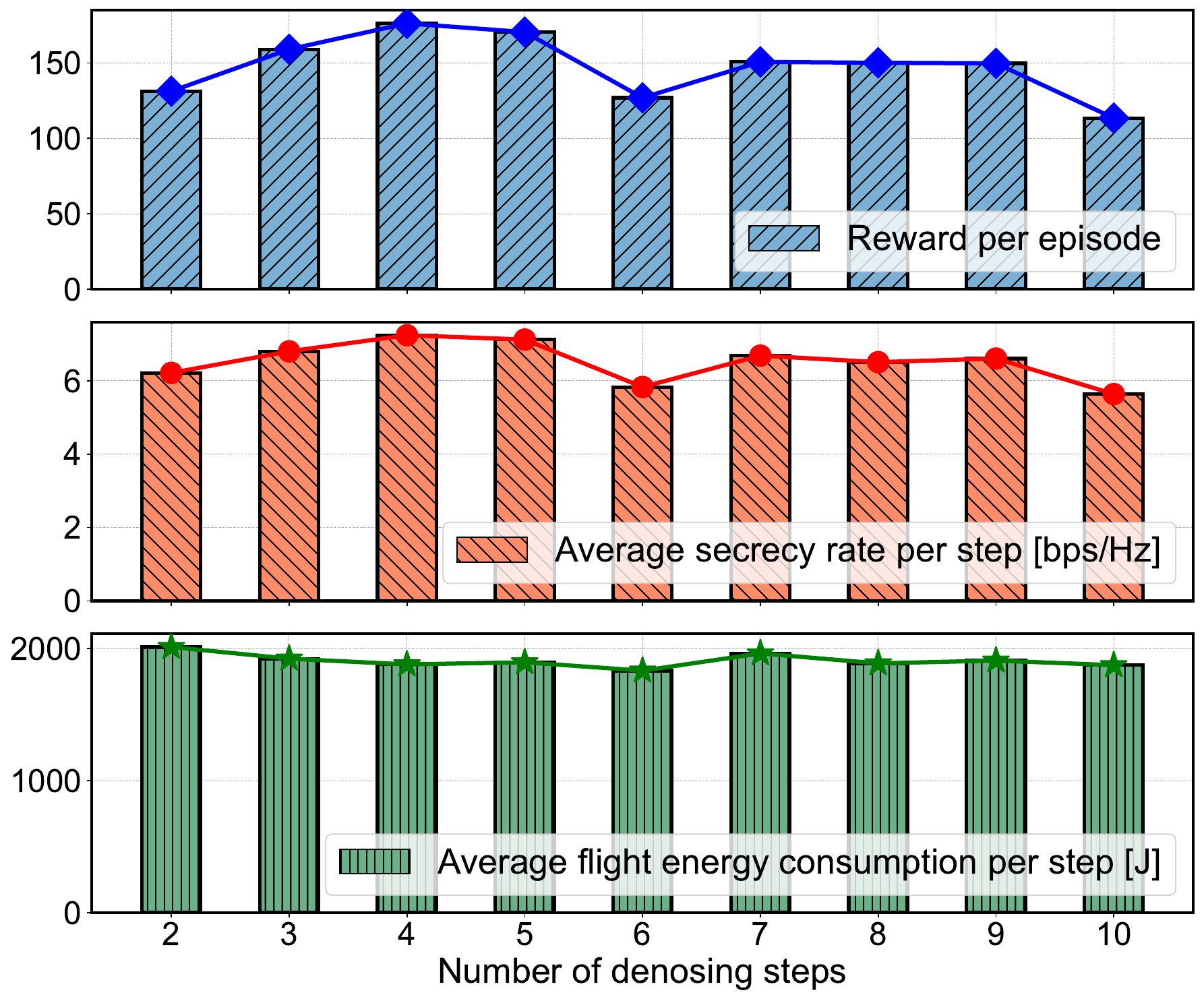}
	\caption{Comparison of curves of GDMTD3 with different denoising steps.}
	\label{Figure: Comparison of curves of GDMTD3 with different denoising steps}
\end{figure}

 \textit{\textbf{Effect of Different Denoising Steps.}} The number of denoising steps in the diffusion reverse process is another critical factor that can significantly impact the performance of diffusion-based models. First, denoising steps determine how effectively the model can reduce noise and generate high-quality samples \cite{Du2024}. Second, an increase in denoising steps also leads to longer training time. Therefore, we compare the impact of varying the number of denoising steps on the performance of GDMTD3. As shown in Fig.~\ref{Figure: Comparison of curves of GDMTD3 with different denoising steps}, increasing the number of denoising steps generally improves the performance of the diffusion model by enabling more precise noise reduction. However, beyond a certain step, which is 4 in the context of our formulated ASCEE-MOP, the benefits of additional denoising steps diminish. This is because increasing the denoising steps can cause the model to overfit the noise pattern. As a result, unnecessary details appear in the generated actions, reducing their quality. The result demonstrates the importance of selecting an appropriate number of denoising steps to balance performance and computational efficiency in the specific problem.

\subsubsection{Impact of Number of UAVs}
\label{SubSubSection: Impact of Number of UAVs}

\par
\begin{figure}[t]
	\centering	\includegraphics[width=\linewidth,scale=1.00]{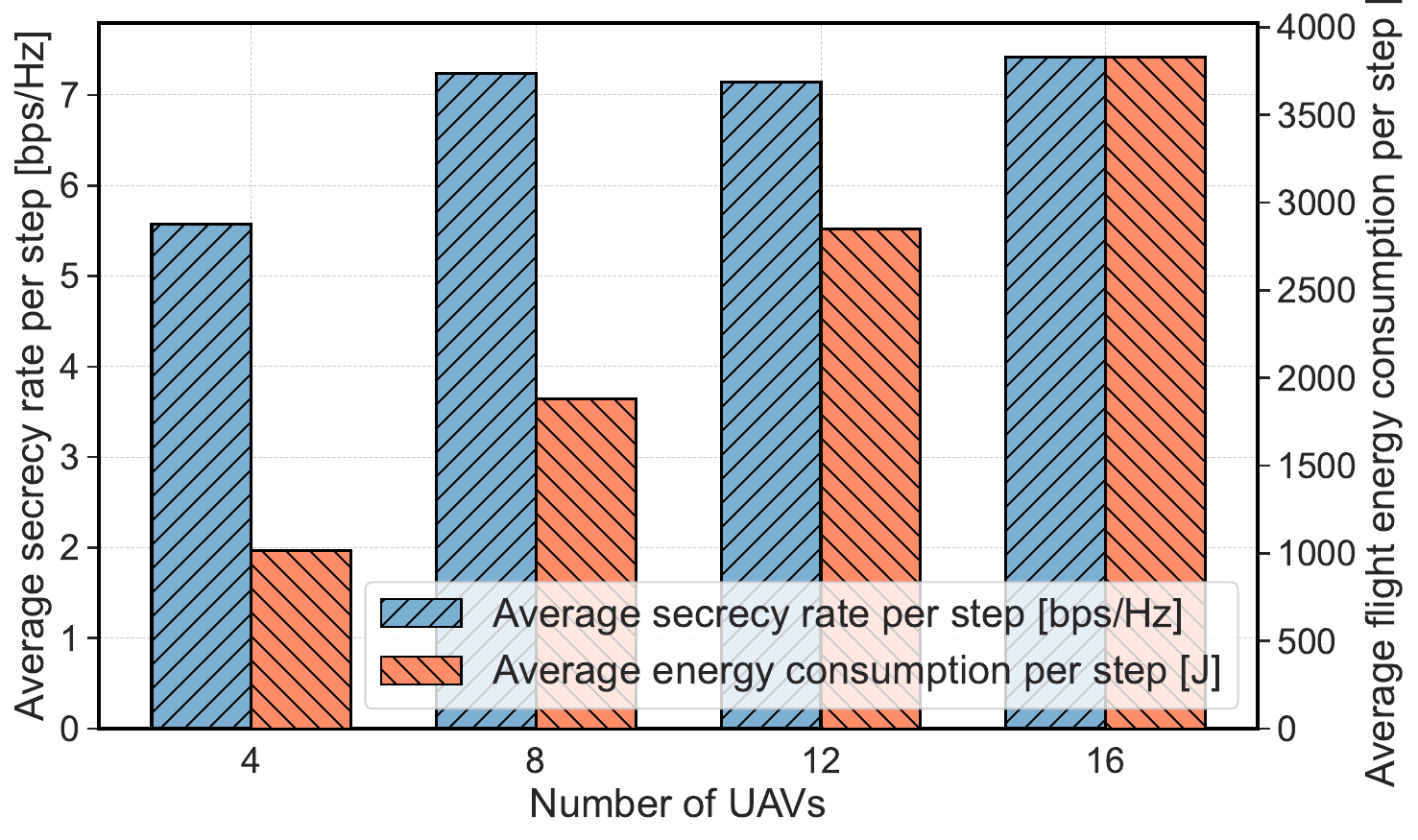}
	\caption{Comparison of curves of GDMTD3 with different UAV numbers.}
	\label{Figure: Comparison of curves of GDMTD3 with different UAV number}
\end{figure}

 To verify the impact of the number of UAVs on system performance, we performed a detailed simulation under varying numbers of UAVs. As shown in Fig.~\ref{Figure: Comparison of curves of GDMTD3 with different UAV number}, the average secrecy rate of the system improves significantly with the initial increase in the number of UAVs. Specifically, when the number of UAVs increases from 4 to 8, the average secrecy rate per step rises from $5.58$ bps/Hz to approximately $7.24$ bps/Hz. This improvement is mainly attributed to the more accurate CB capabilities provided by the denser UAV network. However, the increase in the number of UAVs also leads to higher overall flight energy consumption. For instance, when the number of UAVs increases from $8$ to $16$, the average flight energy consumption per step of the system rises from approximately $1879.85$ J to $2850.38$ J. Moreover, we can notice that after the number of UAVs reaches a certain threshold, the improvement in terms of secrecy rate tends to saturate, while energy consumption still continues to increase. This may be because as the density of UAVs in the fixed space increases, the distance between array elements decreases, potentially leading to increased mutual coupling and interference among UAVs. Consequently, adding more UAVs beyond this number does not significantly enhance the security performance of the system.

\section{Conclusion}
\label{Section:Conclusion}

\par In this work, we investigated a novel UAV swarm-enabled secure surveillance network system, where a UAV swarm perform CB to enhance the security performance between UAV swarm and RBS so as to resist eavesdropping attacks from mobile eavesdroppers. Moreover, we formulated an ASCEE-MOP with an aim to maximize the secrecy rate of the system while minimizing the flight energy consumption of the UAV swarm by optimizing both the excitation current weights and positions of UAVs in conjunction. To solve the non-convex, NP-hard and dynamic optimization problem, we introduced GDMTD3, which effectively captures the high-dimensional probabilistic distributions required for optimal policy decisions. Simulation results demonstrated that the GDMDRL approach outperforms various deployment policies in terms of both the secrecy rate of the system and the flight energy consumption of the UAV swarm. Additionally, the results highlighted the superiority of the GDMTD3 algorithm over several advanced DRL benchmarks in solving the formulated ASCEE-MOP.

\bibliographystyle{IEEEtran}
\bibliography{cite}

\begin{IEEEbiography}[{\includegraphics[width=1in,height=1.25in,clip,keepaspectratio]{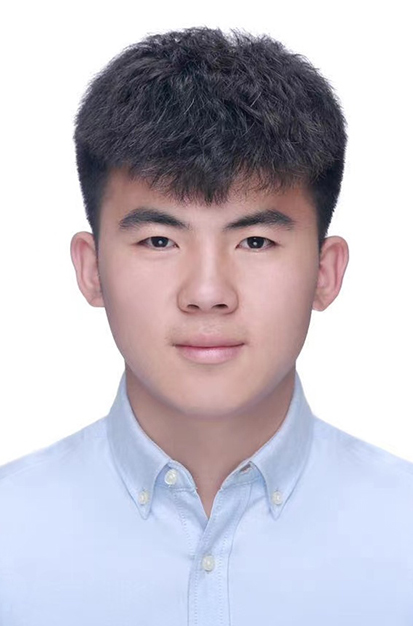}}]{Chuang Zhang} received the B.S. degree in computer science and technology from Jilin University, Changchun, China, in 2021, where he is currently pursuing the Ph.D. degree with the College of Computer Science and Technology. His current research interests include UAV communications, secure communications, distributed beamforming and multi-objective optimization.
\end{IEEEbiography}

\begin{IEEEbiography}[{\includegraphics[width=1in,height=1.25in,clip,keepaspectratio]{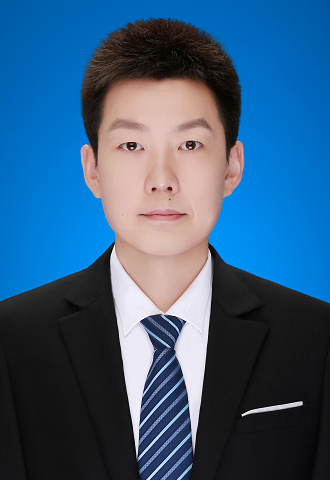}}]{Geng Sun} (Senior Member, IEEE) received the B.S. degree in communication engineering from Dalian Polytechnic University, and the Ph.D. degree in computer science and technology from Jilin University, in 2011 and 2018, respectively. He was a Visiting Researcher with the School of Electrical and Computer Engineering, Georgia Institute of Technology, USA. He is an Associate Professor in College of Computer Science and Technology at Jilin University, and His research interests include wireless networks, UAV communications, collaborative beamforming and optimizations.
\end{IEEEbiography}

\begin{IEEEbiography}[{\includegraphics[width=1in,height=1.25in,clip,keepaspectratio]{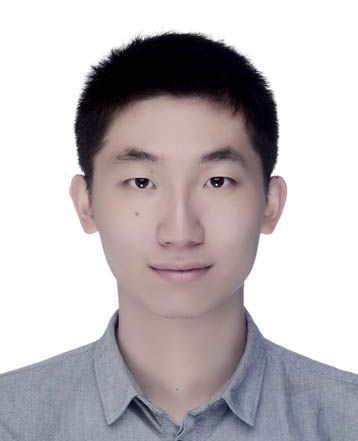}}]{Jiahui Li} received a BS degree in Software Engineering, and an MS degree in Computer Science and Technology from Jilin University, Changchun, China, in 2018 and 2021, respectively. He is currently studying Computer Science at Jilin University to get a Ph.D. degree. His current research focuses on UAV networks, antenna arrays, and optimization.
\end{IEEEbiography}

\begin{IEEEbiography}[{\includegraphics[width=1in,height=1.25in,clip,keepaspectratio]{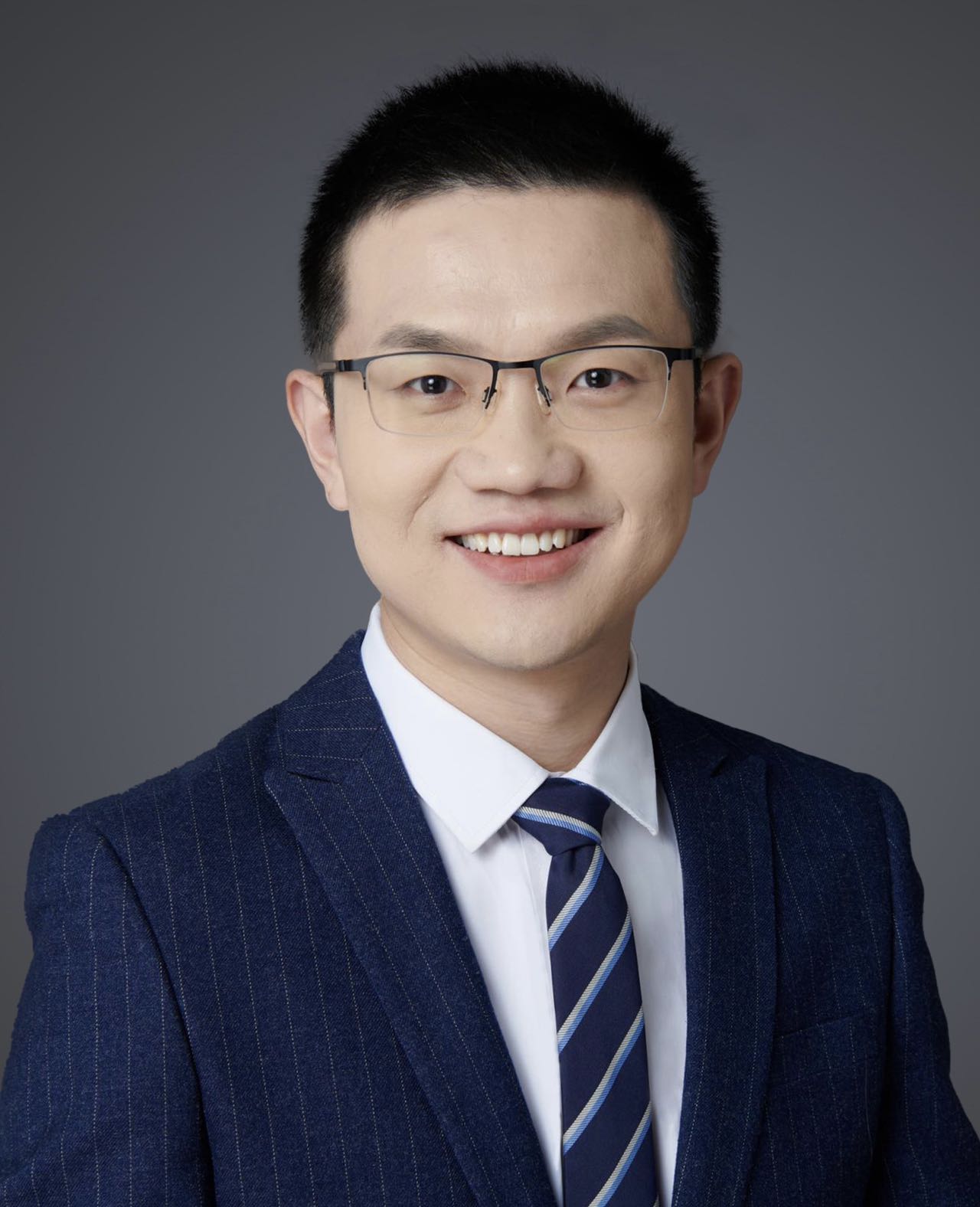}}]{Qingqing Wu} (Senior Member, IEEE) received the B.Eng. and the Ph.D. degrees in Electronic Engineering from South China University of Technology and Shanghai Jiao Tong University (SJTU) in 2012 and 2016, respectively. From 2016 to 2020, he was a Research Fellow in the Department of Electrical and Computer Engineering at National University of Singapore. He is currently an Associate Professor with Shanghai Jiao Tong University. His current research interest includes intelligent reflecting surface (IRS), unmanned aerial vehicle (UAV) communications, and MIMO transceiver design. He has coauthored more than 100 IEEE journal papers with 26 ESI highly cited papers and 8 ESI hot papers, which have received more than 30,000 Google citations. He was listed as the Clarivate ESI Highly Cited Researcher in 2022 and 2021, the Most Influential Scholar Award in AI-2000 by Aminer in 2021 and World’s Top 2\% Scientist by Stanford University in 2020 and 2021.
\par He was the recipient of the IEEE Communications Society Asia Pacific Best Young Researcher Award and Outstanding Paper Award in 2022, the IEEE Communications Society Young Author Best Paper Award in 2021, the Outstanding Ph.D. Thesis Award of China Institute of Communications in 2017, the Outstanding Ph.D. Thesis Funding in SJTU in 2016, the IEEE ICCC Best Paper Award in 2021, and IEEE WCSP Best Paper Award in 2015. He was the Exemplary Editor of IEEE Communications Letters in 2019 and the Exemplary Reviewer of several IEEE journals. He serves as an Associate Editor for IEEE Transactions on Communications, IEEE Communications Letters, IEEE Wireless Communications Letters, IEEE Open Journal of Communications Society (OJ COMS), and IEEE Open Journal of Vehicular Technology (OJVT). He is the Lead Guest Editor for IEEE Journal on Selected Areas in Communications on “UAV Communications in 5G and Beyond Networks”, and the Guest Editor for IEEE OJVT on “6G Intelligent Communications” and IEEE OJ-COMS on “Reconfigurable Intelligent Surface-Based Communications for 6G Wireless Networks”. He is the workshop co-chair for IEEE ICC 2019-2022 workshop on “Integrating UAVs into 5G and Beyond”, and the workshop co-chair for IEEE GLOBECOM 2020 and ICC 2021 workshop on “Reconfigurable Intelligent Surfaces for Wireless Communication for Beyond 5G”. He serves as the Workshops and Symposia Officer of Reconfigurable Intelligent Surfaces Emerging Technology Initiative and Research Blog Officer of Aerial Communications Emerging Technology Initiative. He is the IEEE Communications Society Young Professional Chair in Asia Pacific Region.
\end{IEEEbiography}

\begin{IEEEbiography}[{\includegraphics[width=1in,height=1.25in,clip,keepaspectratio]{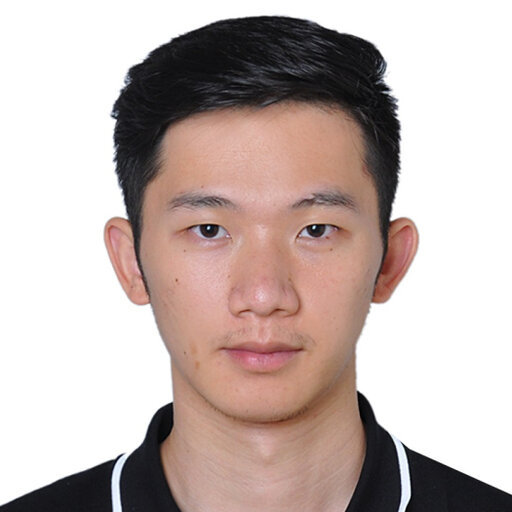}}]{Jiacheng Wang} is the research fellow in the College of Computing and Data Science at Nanyang Technological University, Singapore. Prior to that, he received the Ph.D. degree in School of Communications and Information Engineering, Chongqing University of Posts and Telecommunications, Chongqing, China. His research interests include wireless sensing, semantic communications, and generative AI, Metaverse.
\end{IEEEbiography}

\begin{IEEEbiography}[{\includegraphics[width=1in,height=1.25in,clip,keepaspectratio]{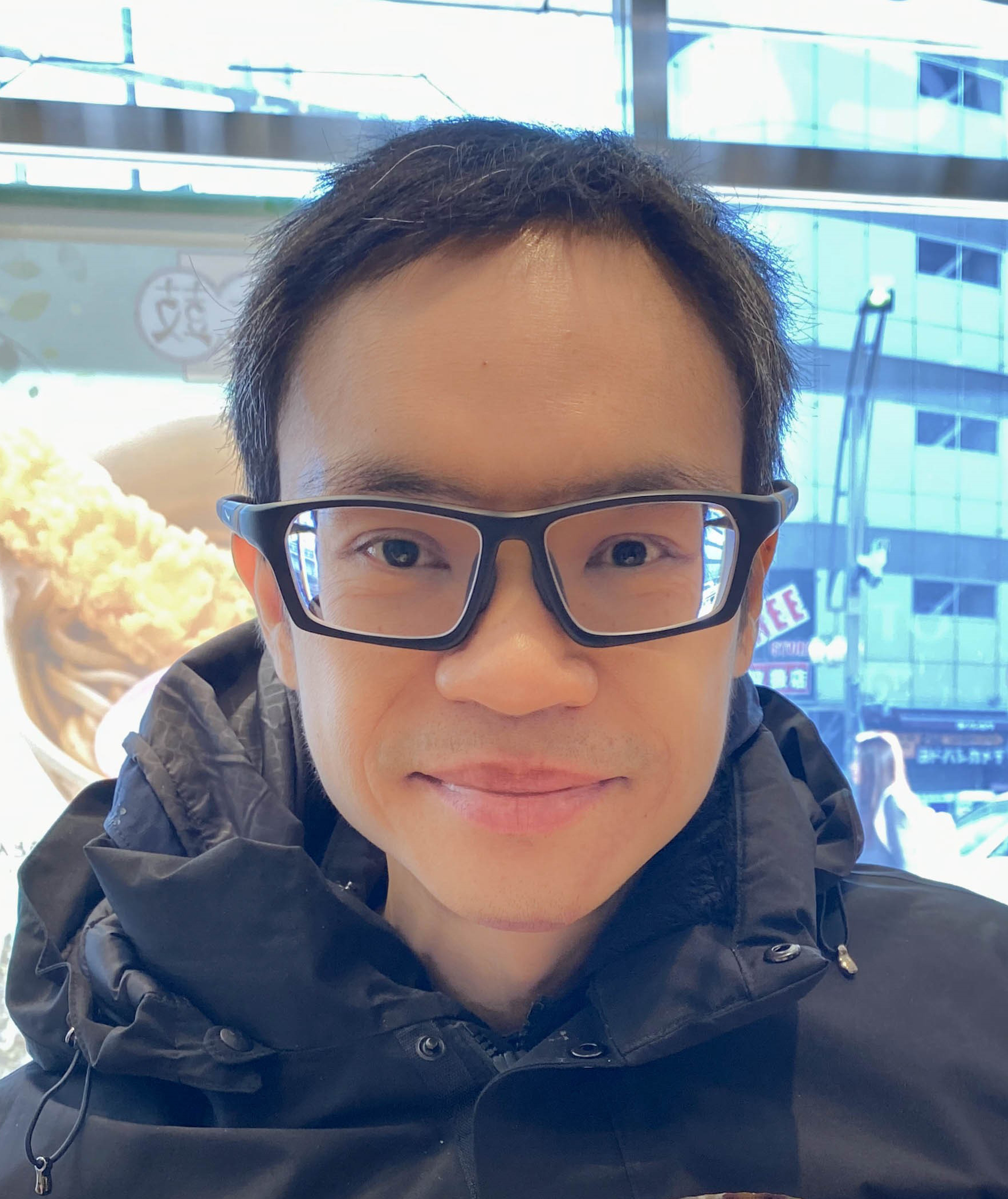}}]{Dusit Niyato} (Fellow, IEEE) received the B.Eng. degree from the King Mongkuts Institute of Technology Ladkrabang (KMITL), Thailand, in 1999, and the Ph.D. degree in electrical and computer engineering from the University of Manitoba, Canada, in 2008. He is currently a Professor with the College of Computing and Data Science, Nanyang Technological University, Singapore. His research interests include the Internet of Things (IoT), machine learning, and incentive mechanism design.
\end{IEEEbiography}

\begin{IEEEbiography}[{\includegraphics[width=1in,height=1.25in,clip,keepaspectratio]{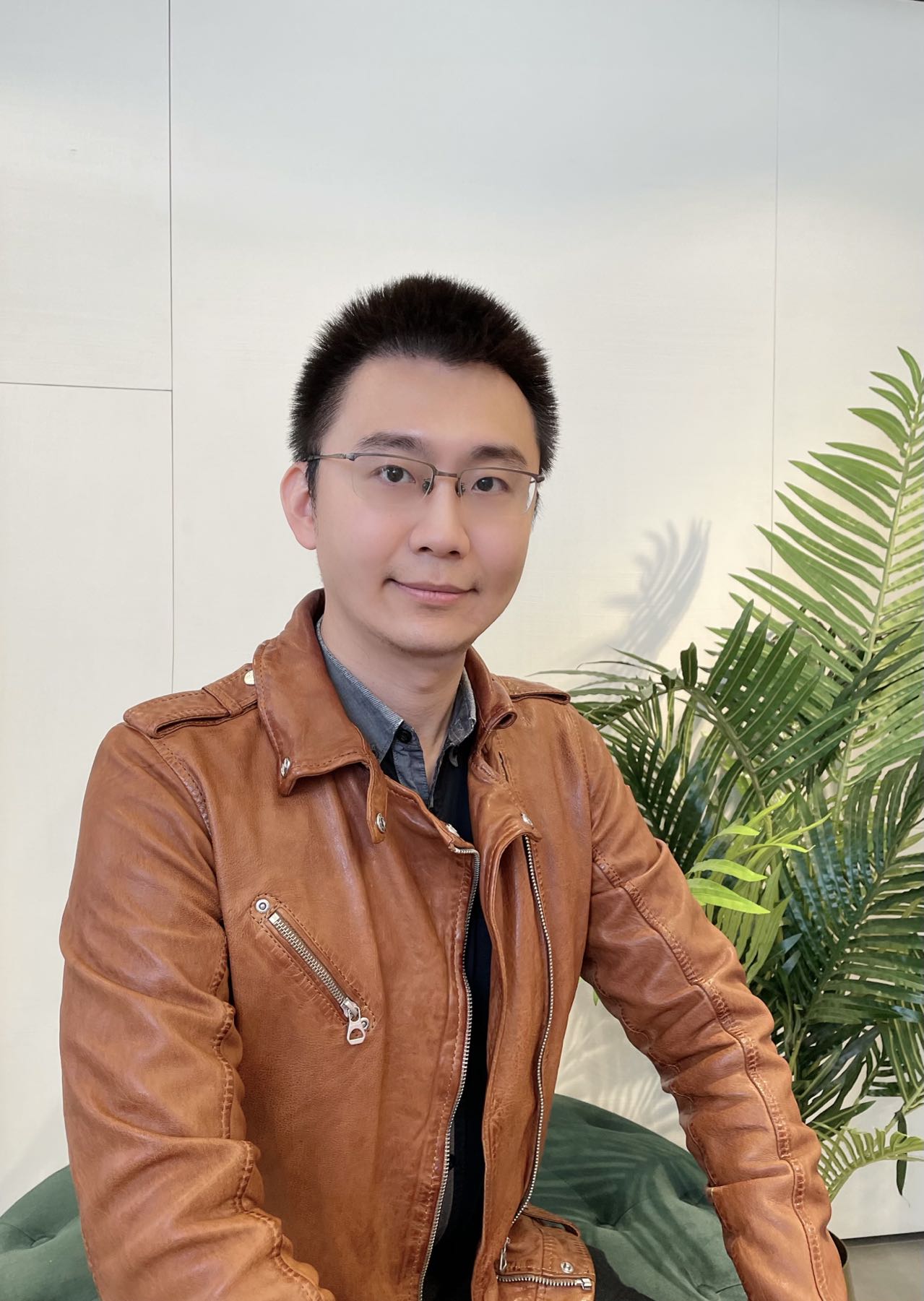}}]{Yuanwei Liu} (Fellow, IEEE) received the PhD degree in electrical engineering from the Queen Mary University of London, U.K., in 2016. He was with the Department of Informatics, King’s College London, from 2016 to 2017, where he was a Post-Doctoral Research Fellow. He has been a Senior Lecturer (Associate Professor) with the School of Electronic Engineering and Computer Science, Queen Mary University of London, since Aug. 2021, where he was a Lecturer (Assistant Professor) from 2017 to 2021. His research interests include non-orthogonal multiple access, reconfigurable intelligent surface, near field communications, integrated sensing and communications, and machine learning.

Yuanwei Liu is a Fellow of the IEEE, a Fellow of AAIA, a Web of Science Highly Cited Researcher, an IEEE Communication Society Distinguished Lecturer, an IEEE Vehicular Technology Society Distinguished Lecturer, the rapporteur of ETSI Industry Specification Group on Reconfigurable Intelligent Surfaces on work item of “Multi-functional Reconfigurable Intelligent Surfaces (RIS): Modelling, Optimisation, and Operation”, and the UK representative for the URSI Commission C on "Radio communication Systems and Signal Processing". He was listed as one of 35 Innovators Under 35 China in 2022 by MIT Technology Review. He received IEEE ComSoc Outstanding Young Researcher Award for EMEA in 2020. He received the 2020 IEEE Signal Processing and Computing for Communications (SPCC) Technical Committee Early Achievement Award, IEEE Communication Theory Technical Committee (CTTC) 2021 Early Achievement Award. He received IEEE ComSoc Outstanding Nominee for Best Young Professionals Award in 2021. He is the co-recipient of the Best Student Paper Award in IEEE VTC2022-Fall, the Best Paper Award in ISWCS 2022, the 2022 IEEE SPCC-TC Best Paper Award, the 2023 IEEE ICCT Best Paper Award, and the 2023 IEEE ISAP Best Emerging Technologies Paper Award. He serves as the Co-Editor-in-Chief of IEEE ComSoc TC Newsletter, an Area Editor of IEEE Communications Letters, an Editor of IEEE Communications Surveys \& Tutorials, IEEE Transactions on Wireless Communications, IEEE Transactions on Vehicular Technology, IEEE Transactions on Network Science and Engineering, and IEEE Transactions on Communications (2018-2023). He serves as the (leading) Guest Editor for Proceedings of the IEEE on Next Generation Multiple Access, IEEE JSAC on Next Generation Multiple Access, IEEE JSTSP on Intelligent Signal Processing and Learning for Next Generation Multiple Access, and IEEE Network on Next Generation Multiple Access for 6G. He serves as the Publicity Co-Chair for IEEE VTC 2019 Fall, the Panel Co-Chair for IEEE WCNC 2024, Symposium Co-Chair for  several flagship conferences such as IEEE GLOBECOM, ICC and VTC. He serves the academic Chair for the Next Generation Multiple Access Emerging Technology Initiative, vice chair of SPCC and Technical Committee on Cognitive Networks (TCCN).
\end{IEEEbiography}

\vfill

\end{document}